\documentclass[twocolumn,           
               showpacs,            
               preprintnumbers,     
               aps,                 
               prd,          	    
               letterpaper,             
               superscriptaddress,      
               nofootinbib,         
               tightenlines,        
               floats,floatfix      
               ]{revtex4-1}

\usepackage{graphicx}
\usepackage{latexsym}
\usepackage{amsmath,amssymb}        
\usepackage[draft=false]{hyperref}

\begin{document}


\title{Bosonic gas as a Galactic Dark Matter Halo}
\author{L. Arturo Ure\~na-L\'opez}%
\email{lurena@fisica.ugto.mx}
\affiliation{Departamento de F\'isica, Divisi\'on de Ciencias e
  Ingenier\'ias, Campus Le\'on, Universidad de Guanajuato, C.P. 37150,
  Le\'on, Guanajuato, M\'exico}%

\author{Argelia Bernal}
\email{abernal@aei.mpg.de}
\affiliation{Max-Planck-Institut f\"ur Gravitationsphysik 
(Albert-Einstein-Institut), Am M\"uhlenberg 1, 14476 Potsdam, Germany}

\date{\today}


\begin{abstract}
We study in detail the properties of gravitationally-bounded
multi-state configurations, made of spin-zero bosons, in the Newtonian
regime. We show that the properties of such configurations, in
particular their stability, depend upon how the particles are
distributed in the different states they are composed of. Numerical
techniques are used to distinguish between stable and unstable
solutions, and to determine the final configurations they evolve
towards to. Multi-state equilibrium configurations can be used as
models of galactic halos made of scalar field dark matter, whose
rotation curves appear more realistic than in the case of single-state
configurations.
\end{abstract}

\maketitle


\section{Introduction \label{sec:introduction}}
It has been known for a long time that, within the context of
Einstein's General Relativity, the luminous matter content of galaxies
cannot explain the so-called Rotational Curves
(RC)\cite{Rubin1983,Persic:1995ru,Primack:2001ia}, which are still
considered one of the cornerstone evidence for the existence of
non-baryonic dark matter. There are many candidates for dark matter
particles, the most popular ones are known as weak interactive massive particles
(WIMPS)\cite{Colafrancesco:2010zn,DelPopolo:2008mr,Taoso:2007qk}. The
accepted paradigm that describes the way in which those particles form
structures is the so called Lambda Cold Dark Matter ($\Lambda$CDM)
model\cite{Olive:2010mh,Benson:2010de,Larson:2010gs}.

An interesting alternative some of us have been working on is to
consider a (real) scalar field as a dark matter candidate, a
hypothesis that has been widely explored in the specialized literature
by many other
authors\cite{Sin:1992bg,Ji:1994xh,Lesgourgues:2002hk,Arbey:2003sj,Matos:2008ag,Matos:2000ng,Matos:2000ss,Lee:1995af,Lee:2008ab,Lee:2008ux,Lee:2008mq,Hu:2000ke,Hu:1998kj,Woo:2008nn,UrenaLopez:2008zh,Lundgren:2010sp};
see also\cite{Matos:2004rs} for a comprehensive review. In most
scalar field models, the dark matter particle is an ultra-light
massive boson, with a Compton wavelength of astrophysical proportions
and a very large mean number density, so that their collective
behavior is well described by a classical scalar field. The Scalar
Field Dark Matter (SFDM) model, as we shall call it in general, offers
the same results as the $\Lambda$CDM model at large scales, up to
linear order
perturbations\cite{Sahni:1999qe,Matos:2000ss,Matos:2009hf,Matos:2008ag,Woo:2008nn,Lundgren:2010sp}.

The RC problem has also been addressed using scalar fields, see for
instance\cite{Ji:1994xh,Sin:1992bg,Lee:1995af,Schunck:1998nq,Lesgourgues:2002hk}). These
works considered the dark matter halo as a Newtonian Bose-Einstein
Condensate (BEC), in which the scalar field dynamics is driven by the
so-called Schroedinger-Poisson system of equations. However, none of
the studies carried on so far have not shown, undeniably, that these
scalar field models can account for all features of realistic galactic
halo.  

The modeling of scalar field halos was based on the (nodeless) ground
state solutions of the SP system, which is the only stable solution,
and the predicted rotation curves are marginally in agreement with the
observed
ones\cite{Ji:1994xh,Guzman:2006yc,Lee:1988av,Guzman:2004wj,Lesgourgues:2002hk,Arbey:2003sj}. In all
cases above, the only stable scalar field configuration is that in
which boson particles are all in the ground state. The ground state is
the only stable solution of the SP system against gravitational
perturbations; other excited configurations are intrinsically
unstable\cite{Guzman:2004wj,Lee:2008ab} (see
also\cite{Schunck:2003kk,Jetzer:1991jr} and references therein).

The main purpose in this work is to further explore the proposal that
was first put forward by Matos \& Ure\~na-L\'opez in
Ref.~\cite{Matos:2007zza}: that realistic scalar field galaxy halos must be
comprised of \emph{multi-state configurations}. As we shall show,
equilibrium configurations of the SP system can be constructed in which
many-particle states coexist simultaneously, so that the whole system
is \emph{stable} under small (radial) perturbations\footnote{The
  relativistic version of the multi-state hypothesis was studied
  recently in\cite{Bernal:2009zy,Bernal:2010zz}, in which stability was also
  confirmed.}. Probably not surprisingly, we have found that RC could
be better fitted by these many-particle systems.

The plan of the paper is as follows. In Sec.~\ref{sec:math}, we
present the mathematical theory behind multi-particle states. In
Sec.~\ref{sec:configurations} we show that their general properties
depend upon the distribution of the particles in the different excited
states. In Sec.~\ref{sec:stability}, we give numerical evidence that
there are stable configurations under small radial perturbations, and
investigate the late time behavior of unstable configurations. The RC
curves predicted by stable multi-state configurations are calculated in 
Sec.~\ref{sec:RotCurves}. Finally, some conclusions are given in
Sec.~\ref{Conclusions}.

\section{Mathematical Background}
\label{sec:math}
Here we give a brief description of a gravitational bounded system of
self-gravitating scalar field particles following the argumentation in
the seminal paper\cite{Ruffini:1969qy}, see
also\cite{Jetzer:1991jr,Schunck:2003kk,Seidel:1990jh,Balakrishna:1997ej,Balakrishna:2006ru}. We
pay special attention to the key aspects needed to build systems that
have particles in the ground state but also in the excited
states\cite{Matos:2007zza}.

We start by assuming a spherically symmetric metric of the
form
\begin{equation}
  ds^2 = - \alpha^2(t,r) dt^2 + a^2(t,r) dr^2 + r^2 d\Omega \,
  , \label{eq:metric}
\end{equation}
in units of $\hbar = c = 1$. The many boson-system is then described
by a (secondly quantized) real scalar field operator of the
form\cite{Birrell:1982ix}
\begin{equation}
\label{quantizedphi}
\hat \Phi = \sum_{nlm} \left[ \hat{b}_{nlm} \Phi_{nlm}(t,{\bf x}) +
\hat{b}_{nlm}^\dagger \Phi_{nlm}^* (t,{\bf x}) \right] \, ,
\end{equation}
where $\hat b_{nlm}$ and $\hat b_{nlm}^\dagger$ are usual annihilation
and creation quantum operators, which obey the commutation relations
\begin{subequations}
  \label{conmutators}
  \begin{eqnarray}
    \left[ \hat{b}_{nlm}, \hat{b}^\dagger_{n'l'm'} \right] &=&
    \delta_{nn'} \delta_{ll'}\delta_{mm'} \, , \\
    \left[ \hat{b}_{nlm}, \hat{b}_{nlm}\right] &=& \left[
      \hat{b}^\dagger_{nlm},\hat{b}^\dagger_{n'l'm'} \right] = 0 \, .
  \end{eqnarray} 
\end{subequations} 
The field coefficients $\Phi_{nlm}$ satisfy the Klein Gordon (KG)
equation in a curved spacetime
\begin{equation}
  \left( \Box-\mu^2 \right) \Phi_{nlm} (t,{\bf x}) = 0 \,
  , \label{eq:KG}
\end{equation}
with $\Box = (1/\sqrt{-g}) \partial_\mu (\sqrt{-g} \, \partial^\mu)$
the covariant d'Alembertian operator, and $\mu$ is the mass of the
bosons. The most general solution of Eq.~(\ref{eq:KG}) is of the form
\begin{equation}
  \Phi_{nlm} (t,{\bf x}) = R_{nl} (t,r) Y_{lm}(\theta,\varphi) \,
  , \label{eq:eigenphi}
\end{equation}
where we $R_{nl}(t,r)$ is the radial function to be determined from
the KG equation. The scalar product of the functions above is defined
as
\begin{equation}
  (\Phi_{nlm},\Phi_{n^\prime l^\prime m^\prime}) \equiv -i \int_\Sigma
  \Phi_{nlm} \partial_\mu \Phi^*_{n^\prime l^\prime m^\prime} \, n^\mu
  \, \alpha \sqrt{\gamma} \, d\Sigma \, , \label{eq:s-product}
\end{equation}
where $\gamma$ is the determinant of the $3$-dim metric on the
spacelike hypersurface $\Sigma$, $n^\mu$  is a timelike unit vector
orthogonal to $\Sigma$, and $d\Sigma$ is the volume element. In our
case, see Eqs.~(\ref{eq:metric}) and~(\ref{eq:eigenphi}),
Eq.~(\ref{eq:s-product}) reads
\begin{equation}
  (\Phi_{nlm},\Phi_{n^\prime l^\prime m^\prime}) = -i \, \delta_{l
    l^\prime} \delta_{m m^\prime} \int_V R_{nl} \, \partial_t \,
  R^*_{n^\prime l^\prime} \, a^2(t,r) r^2 \, dr \, , 
\end{equation}
where we have made use of the orthogonality condition of the spherical
harmonics $Y_{lm}$.

Assuming that there exists a vacuum state defined by
\begin{equation}
  \hat{b}_{nlm} \, |0, 0, \ldots, 0 \rangle = 0 \, , \quad \forall \,
  (n,l,m)
\end{equation}
we can construct the orthonormal many-particle states
\begin{eqnarray}
  \label{state}
  |Q\rangle &=& |N_{100},N_{200},N_{21-1},N_{210}, \ldots \rangle \, ,
  \\
  &\equiv& \frac{(\hat{b}^\dagger)^{N_{100}}
    (\hat{b}^\dagger)^{N_{200}} \cdots}{N_{100}! N_{200}! \cdots}
  |0,0, \ldots 0 \rangle \, , \nonumber
\end{eqnarray}
composed of many scalar particles distributed in sets of 
${}^iN_{nlm}$ particles of mass $\mu$ with angular momentum $l$ and
azimuthal momentum $m$; the $n$ sub-index labels the eigenstates
according to their radial function $R_{nl}$. Notice that many-particle
states are constructed from the vacuum through the repeated
application of the creation operators $\hat{b}^\dagger$.

On the other hand, the gravitational field is a classical field whose
dynamics is described by the Einstein equations
\begin{equation}
G_{\alpha \beta} = 8\pi G \langle Q| :\hat{T}_{\alpha\beta}: |Q \rangle
\, , \label{eq:einsteineq}
\end{equation}
where the source on the r.h.s. is the expectation value of the
energy-momentum tensor operator
\begin{equation}
  \hat{T}_{\alpha\beta} = \partial_{\alpha} \hat{\Phi} \partial_{\beta}
  \hat{\Phi} - \frac{1}{2} g_{\alpha \beta} \left( \partial^\sigma
  \hat{\Phi} \partial_\sigma \hat{\Phi} + \mu^2 \hat{\Phi}^2 \right)
  \, . \label{eq:quantizedt}
\end{equation}
Notice that we are implicitly assuming the so called normal ordering
operation, $:\hat{T}_{\alpha\beta}:$, so that
\begin{subequations}
  \begin{eqnarray}
    :\hat{b}_{nlm} \hat{b}^\dagger_{nlm}: = \hat{b}^\dagger_{nlm}
    \hat{b}_{nlm} \, , \\
    \langle 0, \ldots, 0, 0| :\hat{T}_{\alpha\beta}: |0, 0, \ldots, 0
    \rangle = 0 \, ,
  \end{eqnarray}
\end{subequations}
and then the (divergent) vacuum energy density identically vanishes.

The orthogonality of the quantum states ensures that the expectation
value is given as a superposition of the expectation values of the
energy-momentum tensor components for each individual state, that is,
\begin{equation}
  \langle Q| :\hat{T}_{\alpha\beta}: |Q \rangle = \sum^\infty_{n=1}
  \sum^{n-1}_{l=1} \sum^{l}_{m=-l} \langle N_{nlm} |
  :\hat{T}_{\alpha \beta}: | N_{nlm} \rangle \, , 
\end{equation}
where we are defining the \emph{single} states
\begin{equation}
  | N_{nlm} \rangle \equiv |0,0,\ldots,N_{nlm},0,\ldots,0 \rangle \,
  . \label{eq:single}
\end{equation}
Hence, the Einstein equations~(\ref{eq:einsteineq}) read
\begin{equation}
G_{\alpha \beta} = 8\pi G \sum_{n,l,m} T_{\alpha \beta (nlm)}
\, , \label{eq:einsteinequation}
\end{equation}
where
\begin{eqnarray}
  T_{\alpha\beta (nlm)} &=& \partial_{\alpha} \Phi_{nlm}
  \partial_{\beta} \Phi^*_{nlm} - \frac{1}{2} g_{\alpha \beta} \left(
  \partial^{\sigma} \Phi_{nlm} \partial_{\sigma} \Phi^*_{nlm}
  \right. \nonumber \\
  && \left. + \mu^2 \Phi_{nlm} \Phi^*_{nlm} \right) \, ,
\end{eqnarray}
and we also normalized the eigenfunctions so that $\Phi \to
\Phi/\sqrt{2N_{nlm}}$.

Therefore, in the case when particles populate various excited levels,
the source of the Einstein equations~(\ref{eq:einsteinequation}) is
equivalent to the energy momentum tensor of many (independent)
classical complex scalar fields $\Phi_{nlm}(t,{\bf x})$ minimally
coupled to gravity. Each one of such scalar fields accounts for only
one of the excited single states~(\ref{eq:single}), and its dynamics
is given by its own KG
equation~(\ref{eq:KG})\cite{Ruffini:1969qy,Birrell:1982ix}.
 
Finally, we consider the Newtonian limit of the coupled
Einstein-Klein-Gordon (EKG) equations~(\ref{eq:KG})
and~(\ref{eq:einsteinequation}), which results in the so-called
Schr\"odinger-Poisson (SP) system\cite{1987PhRvD..35.3640F}
\begin{subequations}
  \label{eq:SP}
  \begin{eqnarray}
    \nabla^2 U &=& \sum_{nlm}| \Psi_{nlm}|^2 \, , \\
    i\partial_t \Psi_{nlm} &=& -\frac{1}{2} \nabla^2 \Psi_{nlm} +U
    \Psi_{nlm}\, , \label{eq:SP-a}
\end{eqnarray} 
\end{subequations}
where $\Psi_{nlm}$ is related to $\Phi_{nlm}$ by
\begin{equation} 
  \sqrt{8\pi G} \, \Phi_{nlm} (t,{\bf x}) = e^{-i\mu t}
  \Psi_{nlm}(t,{\bf x}) \, .
\end{equation}
Then, the Newtonian  version of the EKG equations describes  the
dynamics of non-relativistic wave functions which are coupled among
themselves through the Newtonian gravitational potential $U$.

Once in the non-relativistic regime, we can define physical quantities
like the kinetic $K$ and gravitational $W$ energies, and the total
number of particles $\mathcal{N}$. These quantities can be explicitly
given in terms of the Newtonian fields as\cite{Guzman:2004wj}
\begin{subequations}
\begin{eqnarray}
K &=& -\frac{1}{2} \sum_{n,l,m} \int(\Psi_{nlm}^* \nabla^2 \Psi_{nlm} +
\Psi_{nlm}\nabla^2 \Psi_{nlm}^* ) dv  \, , \\
W &=& \sum_{n,l,m} \int U |\Psi_{nlm}|^2 dv \, , \\
\mathcal{N} &=& \sum_{n,l,m} \int| \Psi_{nlm}|^2 dv \, .
\end{eqnarray}   
\end{subequations}
These expressions will be useful later to monitor the numerical
evolution of multi-state configurations studied in
Sec.~\ref{sec:stability}.

\section{Mixed Newtonian states}
\label{sec:configurations}
In this section we construct solutions of the SP system~(\ref{eq:SP})
when $\mathcal N$ bosons are allowed to occupy $\mathcal I$ different
levels, all of which, for simplicity in the discussion, will have zero
angular momentum $(l=0,m=0)$. Hence, the states are of the form
$|Q\rangle = |N_1,N_2,N_3,...,N_{\mathcal I} \rangle$. Assuming
spherical symmetry, we then have
\begin{subequations}
  \label{eq:SPspherical}
  \begin{eqnarray}
    \frac{1}{r^2} \frac{\partial^2 (r^2 U)}{\partial r^2} &=&
    \sum^\mathcal{I}_{n=1} |\Psi_{n}|^2 \, , \\
    i\frac{\partial \Psi_{n}}{\partial t} &=& - \frac{1}{2r^2}
    \frac{\partial^2 (r^2 \Psi_n)}{\partial r^2} + U \Psi_{n} \, ,
    \quad n=1,..,\mathcal{I}
\end{eqnarray}
\end{subequations}

First of all, we look for stationary equilibrium configurations in the
form
\begin{equation}
\Psi_{n} = e^{-i\gamma_n t} \phi_n(r) \, , \label{eq:Psi}
\end{equation}
for which the system~(\ref{eq:SPspherical}) becomes
\begin{subequations}
  \label{eq:SPharmonic}
   \begin{eqnarray}
    \frac{1}{r^2} \frac{d^2 (r^2 U)}{d r^2} &=& \sum_{n=1}^{\mathcal
      I} |\phi_n|^2 \, ,\\
    \frac{1}{r^2} \frac{d^2 (r^2 \phi_n)}{d r^2} &=&2 ( U - \gamma_n)
    \phi_{n} \, ,
    \quad n=1,..,\mathcal{I}
  \end{eqnarray}
\end{subequations}
\begin{table*}
\begin{ruledtabular}
  \begin{tabular}{lccccc}
    State& $\phi_n(0)$&$\gamma_n$ & $K$ & $W$ & $\mathcal{N}$ \\
    \hline
    $|N_{1}, 1.1 \rangle$ & $\phi_2(0) =0.756$ & $\gamma_1 = -1.033$,
    $\gamma_2 =-0.574$ & 1.846 & -3.691 & 3.493 \\
    $|N_{1}, 1.6 \rangle$ & $\phi_2(0) =0.934$ & $\gamma_1 =-1.163$,
    $\gamma_2 =-0.677$  & 2.272 & -4.544 & 3.945 \\
    $|N_{1}, 0.96, 0.91 \rangle$ & $\phi_2(0) =0.710$, $\phi_3(0) =0.543$ &
    $\gamma_1 =-1.185$, $\gamma_2 =-0.712$, $\gamma_3 =-0.471$ & 2.407
    & -4.819 & 4.520
\end{tabular}
\end{ruledtabular}
\caption{Central values of excited states $\phi_n(0)$, eigenvalues
  $\gamma_n$, kinetic $K$ and gravitational $W$ energies, and the
  total number of particles $\mathcal{N}$ of three mixed states, all
  of them with $\phi_1(0)=1.0$. The labeling of the mixed states is in
  terms of the $\eta$ parameters defined in Eq.~(\ref{eq:ratios}). See
  text bellow for details.}
\label{table1}
\end{table*}

\begin{figure}[htp]
\includegraphics*[width=6.5cm,angle=-90]{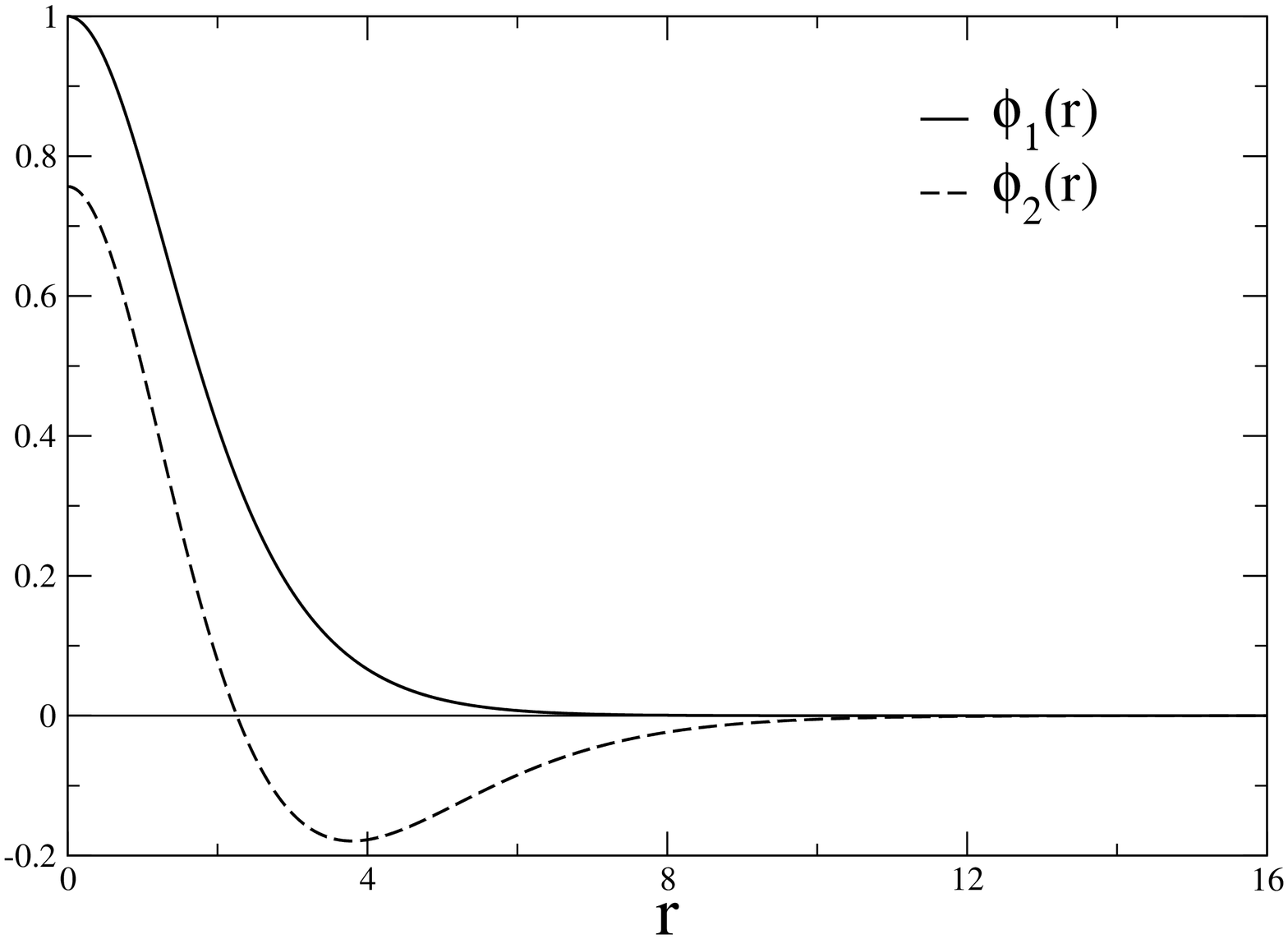}
\includegraphics*[width=7cm]{fig2.eps}
\caption{\label{fig:fig1} Radial functions $\phi_1$ and $\phi_2$
  for the mixed configuration $|N_{1},1.1\rangle$, and $\phi_1$, 
  $\phi_2$ and $\phi_3$ for the mixed configuration 
  $|N_{1},0.96,0.91\rangle$ see also the first and third 
  entries in Table~\ref{table1} and text below for details about its
  construction.}
\end{figure}

Eqs.~(\ref{eq:SPharmonic}) will be solved under the following boundary
conditions. In order to obtain regular solutions at the origin, we
demand that the spatial derivatives of all $\phi_n$ and $U$ must
be zero at this  point, but we arbitrarily prescribe the central
values of all $\phi_n$, as these values are the free parameters of the
solutions. Also, we impose $\phi_n(r) \to 0$ and $U(r)=-\mathcal N/r$
as $r \to \infty$ because we are looking for bounded configurations.

With these boundary conditions the system becomes an eigenvalue
problem. Given the central values \{$\phi_n(0)$\} there are unique
values \{$\gamma_n$\} and $U(0)$ for which the boundary conditions are
satisfied. The numerical solutions are then found by using a shooting
method to integrate Eqs.~(\ref{eq:SPharmonic}) from $r=0$ up to the
numerical boundary $r=r_{max}$, with \{$\gamma_n$\} and $U(0)$ playing
the role of shooting parameters.

Because the equations are integrated in a finite numerical domain, it
is more convenient to introduce more detailed specifications to the
boundary conditions at $r \to \infty$. We use the asymptotic behavior
of the solutions: the gravitation potential goes like $U(r \to \infty)
= -\mathcal{N}/r$, and the radial functions behave as $\phi_n(r \to
\infty) \sim \exp(-\sqrt{-2\gamma_n}r)$. Then, more suitable boundary
conditions are
\begin{subequations}
  \label{eq:bcrmax}
   \begin{eqnarray}
    U(r_{max})+r_{max}U^\prime_{max}=0\,,\\
    \phi^\prime_n(r_{max}) + \sqrt{2\gamma_n^2} \, \phi_n(r_{max})=0 \, .
  \end{eqnarray}
\end{subequations}
The shooting procedure is then used for different values of $r_{max}$. As 
$r_{max}$ is increased the shooting parameters converge, and we choose
as solutions those which satisfy the boundary
conditions~(\ref{eq:bcrmax}) within a prescribed tolerance.

Before we show the numerical results we make some remarks about
notation. In the case when all the particles are in the one same level,
the system (\ref{eq:SPharmonic}) consists of only one Schr\"odinger
equation and only one wave function in the source term of the Poisson
equation. Such system has been widely studied in~\cite{Guzman:2004wj},
where several solutions \{$\phi_n,$ $\gamma_n,$ $U$ \} have been
calculated.
 
The main difference between the functions $\phi_n$ is that they have
($n-1$)-nodes in their radial profile. The state corresponding to the
zero-node function $\phi_1$ has the lowest (negative) total energy
$E=K+W$, the lowest energy eigenvalue $\gamma_1$, and is
correspondingly the less massive; this state is called the ground
state. The other solutions with nodes are more massive and have larger
energy values than the ground state; they are called excited states. 

In this spirit, we shall call mixed states to those configurations in
which single states $\phi_n$ are present simultaneously. For instance, 
the simplest mixed state configuration consists of the ground state 
$\phi_1$ and the first excited state $\phi_2$, and then it is
characterized by quantum state $|N_{1}, N_{2}\rangle$. In the next
section we construct some of these configurations.

\subsection{Ground-first mixed states $|N_{1}, N_{2}\rangle$}
\label{subsec:mixstatescons}
Let us start with the simplest mixed state, that composed of the
ground state $\phi_1$ plus the first excited state $\phi_2$. At this
point, it is not necessary to solve Eqs.~(\ref{eq:SPharmonic}) for
each possible pair $(\phi_1(0),\phi_2(0))$ in order to find the
complete space of solutions. 

Instead, it is worth to use the scaling symmetry that the complete SP
system obeys\cite{Guzman:2004wj}; for the particular case here, it
reads 
   \begin{eqnarray}
    \label{eq:escalingsol}
    \{\phi_1,&\phi_2,&\gamma_1,\gamma_2,U,r\}\to \\
&&\{\lambda^2\hat \phi_1, \lambda^2 \hat
\phi_2,\lambda^2\hat\gamma_1,\lambda^2\hat\gamma_2,\lambda^2\hat
U,\lambda^{-1}r\} \nonumber\, ,
   \end{eqnarray}
where $\lambda$ is an arbitrary parameter. 

This means that once we have found a solution to the SP system for
given values of $(\hat \phi_1(0),\hat \phi_2(0))$, there is a complete
set solutions each of which are related to each other just by the
scaling transformation~(\ref{eq:escalingsol}). This set we will call
it a family of solutions. Different families are then found by taking
different central values of $(\hat \phi_1(0),\hat \phi_2(0))$.

The central values $\phi_1(0)$,$\phi_2(0)$ of the solutions in a
family, as a consequence of the scaling
relation~(\ref{eq:escalingsol}), will be  located along the straight
line defined by the origin and the given point $(\hat \phi_1(0),\hat
\phi_2(0))$ on the plane $\phi_1,\phi_2$. Therefore, once all the
solutions with the same value $\hat \phi_1(0)$ and different
$\hat{\phi}_2(0)$ are known, and their respective families have been
calculated, the complete space of solutions can be constructed as the
collection of all families of solutions. For simplicity in the
notation, we will drop the caret symbol from the field quantities, in
the understanding that we are dealing with scaled quantities.

Useful quantities to characterize mixed states are the ratios of the
number of particles in different states with respect to the ground
state; we define these ratios as
\begin{equation}
  \label{eq:ratios}
  \eta_n \equiv \mathcal{N}_n/\mathcal{N}_1 \, ,
\end{equation}
where by definition $\eta_1 = 1$, and the total number of particles is
$\mathcal{N} = \mathcal{N}_1 (1 + \eta_2 + \eta_3 + \cdots )$. Notice
that $\eta_n$ are invariant quantities under the scaling
relationship~(\ref{eq:escalingsol}).

We calculated solutions with $\phi_1(0)=1$ for different values of
$\phi_2(0)$, and we observed that $\eta_2$ increases monotonically as
$\phi_2(0)$ grows, starting at the value $\eta_2 = 0$ when
$\phi_2(0)=0$ (no particles in the excited state, $N_2=0$). This
behavior of $\eta_2$ allows us to choose it as a free parameter
instead of $\phi_2(0)$, and then $(\phi_1(0),\eta_2)$ will be the
representative parameters in the construction of the solutions. This
is a convenient option because, as we will see in the next section,
there is evidence that the stability under radial perturbations of a
ground-first state depends mainly on the values of $\eta_2$.
 
In order to fix the value of $\eta_2$, it is necessary to add the
differential expressions for the number of particles $\mathcal{N}_j$
\begin{equation}
  \label{eq:difnop}
\frac{d \mathcal{N}_n}{dr}=\phi_j^2 r^2\, , \quad n=1,..,\mathcal{I}
\, ,
\end{equation} 
in the system (\ref{eq:SPharmonic}). The  boundary conditions for
these equations are given at the origin $N_n(0)=0$, and the desired
value of $\phi_2$ is imposed at $r_{max}$. We then solve the system of
equations (\ref{eq:SPharmonic}) and~(\ref{eq:difnop}) using a shooting
method; this time, however, the shooting parameters are $\eta_2(0)$,
$\gamma_1$, $\gamma_2$ and $U(0)$.
 
Once again, the complete space of  configurations is formed by the 
collection of families of solutions, each one characterized by the
same value of $\hat \phi_1(0)$ and different $\eta_2$. As a
consequence of (\ref{eq:escalingsol}), the physical quantities of the solutions for each family will be related by 
   \begin{eqnarray}
    \label{eq:escalingpq}
    \{ N_1,N_2,K,W \}&& \to \\    
    && \{ \lambda \hat{N}_1, \lambda \hat{N}_2,\lambda^3 \hat{K},
    \lambda^3 \hat{W} \} \nonumber \, .
   \end{eqnarray}

In the top panel of fig.~\ref{fig:fig1} we show typical radial 
functions of a (mixed) ground-first state, the zero-node radial
function corresponds to the ground state whereas the one-node radial
function corresponds to the first excited state. The mixed state was 
constructed with $\phi_1(0) =1.0$ and $\eta_2 =1.1$, and so we labeled
it as $|\mathcal{N}_{1}, 1.1 \rangle$. 

In the first row of Table~\ref{table1}, we show the scalar field
central value for the excited state $\phi_2(0)$, the frequencies for
both states, $\gamma_1$ and $\gamma_2$, as well as the kinetic and the
gravitational energies, and the total number of particles of the
system. The same quantities are shown in the second row of the same
table for the mixed state $|\mathcal{N}_{1}, 1.6 \rangle$, which
was constructed with $\phi_1(0) =1.0$ and $\eta_2 =1.6$.

Following a similar procedure to the explained above we also
constructed systems where particles are coexisting in the ground state
and in the first and second excited states. In the bottom panel of
Fig.~\ref{fig:fig1}, we show the radial functions for one of those
systems with $\phi_1(0)=1.0$, and $\eta_2 = \eta_3 =1.0$; we called it
$|\mathcal{N}_1,1.0,1.0 \rangle$. 

In Table~\ref{table1} we show its energy eigenvalues $\gamma_{1,2,3}$
and other physical quantities. We were able to verify that, within the
limitations imposed by the numerical error, all configurations we
could find satisfy the virialization condition $2K +W =0$.

We constructed several ground-first states for different values of
$\phi_1(0)$ and $\eta_2$. In Fig.~\ref{fig:fig4}, we plot the total
energy of each system in terms of the number of particles in the
ground state and first excited states, $N_1,N_2$. It is possible 
to notice that the total energy for each system is negative, which 
implies that they are gravitationally bounded objects. 

In Fig.~\ref{fig:fig5}, the energy eigenfrequencies $\gamma_1$ and
$\gamma_2$ are shown; for all configurations $\gamma_1 <
\gamma_2$. Finally, in Fig.~\ref{fig:fig6}, the kinetic energy of each
state, $K_1$ and $K_2$, is shown. In contrast with the behavior of the
eigenfrequencies, there are configurations for which $K_1 <
K_2$ if $\eta_2 > 1$, whereas $K_1 > K_2$ if $\eta_2 < 1$.

\begin{figure}[t]
\centering
\includegraphics*[width=7cm,angle=-90]{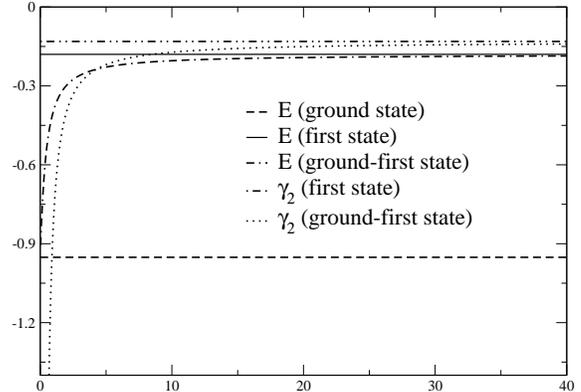}
\caption{\label{fig:fig3} Comparison of the total energy $E = K + W$
  of a ground-first configuration with a fixed number of particles
  $\mathcal{N} = \mathcal{N}_1 + \mathcal{N}_2 =2.0622$, with two
  single-state configurations with the same number of total particles:
  one ground single-state, and one first excited single-state. The
  total energy depends upon the fraction $\eta_2 =
  \mathcal{N}_2/\mathcal{N}_1$, see Eq.~(\ref{eq:ratios}). If
  $\mathcal{N}_1 \gg \mathcal{N}_2 $, $E$ goes to the energy value of
  the ground single-state; if $\mathcal{N}_1 \ll \mathcal{N}_2 $, then
  $E$ goes to the energy value of the first excited single-state. We
  make a similar comparison with $\gamma_2$. If $\mathcal{N}_1 \gg
  \mathcal{N}_2$, $\gamma_2$ takes very large negative values, whereas
  if $\mathcal{N}_1 \ll \mathcal{N}_2$, then $\gamma_2$ approaches the
  value corresponding to a first excited single-state.}
\end{figure}

\subsection{Mixed states as a generalization of single states}
We want to show now that mixed states are generalizations  
of single states. As mentioned before, if the number of particles
$\mathcal{N}$ in an single state is fixed, whatever the ground or any
of the excited states, the total energy $E$ takes a fixed value
because of the eigenvalue problem; this value however increases for
larger number of nodes in the radial profile.

This behavior of $E$ changes radically if the same number of particles
$\mathcal{N}$ are distributed in a mixed configuration. Now, the total
energy of the system $E$ can take values from a continuum interval
depending on how the particles populate the states of the mixed
state. 

In particular, for a ground-first configuration the total energy is a
monotonically increasing function of $\eta_2$. If $\eta \sim 0$, $E
\to E^{(\mathrm{single})}_1$, where $E_1$ is the total energy of a single
ground state composed of $\mathcal{N}$ particles. On the other hand,
if the particles are moved from the ground to the first excited state,
the total energy of the mixed state takes continuum values, and $E \to
E^{(\mathrm{single})}_2$, where $E^{(\mathrm{single})}_2$ is the total
energy of a single first state, as the ground state is depopulated (in
other words, $\eta_2 \to \infty$).

In Fig.~\ref{fig:fig3} we show this behavior of the total energy for a
system of with a fixed total number of particles $\mathcal{N}
=2.0622$. Notice that the values are within the range
$E^{(\mathrm{single})}_1 < E < E^{(\mathrm{single})}_2$. The same kind
of behavior was found for all other characteristic quantities, like
the eigenfrequencies $\gamma_n$, see also Fig.~\ref{fig:fig3}. 

From this perspective, we can say that, given a system of
$\mathcal{N}$ particles in a single-state configuration, it is
possible to change the properties of the quantities attached to it by
moving particles away to populate other excited states.

\begin{figure}[t]
\centering
\includegraphics*[width=7cm]{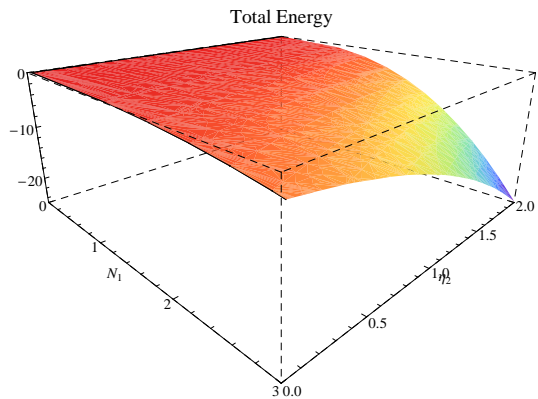}
\caption{\label{fig:fig4}Total energy of ground-first configurations 
in terms of $\mathcal{N}_1$ and $\mathcal{N}_2$. All of them are
negative, which implies that the systems are gravitationally
bounded.}
\end{figure} 

\begin{figure}[t]
\centering
\includegraphics*[width=7cm]{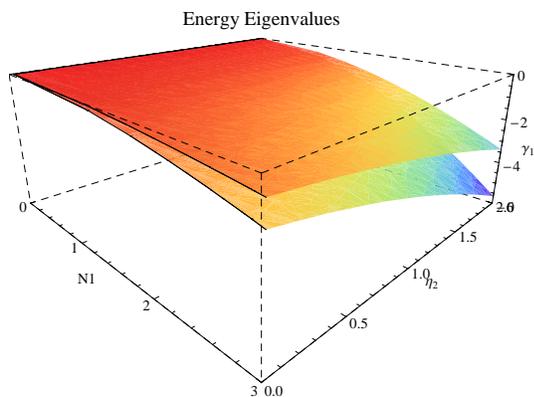}
\caption{\label{fig:fig5} Frequency eigenvalues $\gamma_1$ and
  $\gamma_2$ of ground-first configurations in terms of
  $\mathcal{N}_1$ and $\mathcal{N}_2$. For all the configurations it
  is satisfied that $|\gamma_2| < |\gamma_1|$.}
\end{figure}

\begin{figure}[t]
\centering
\includegraphics*[width=7cm]{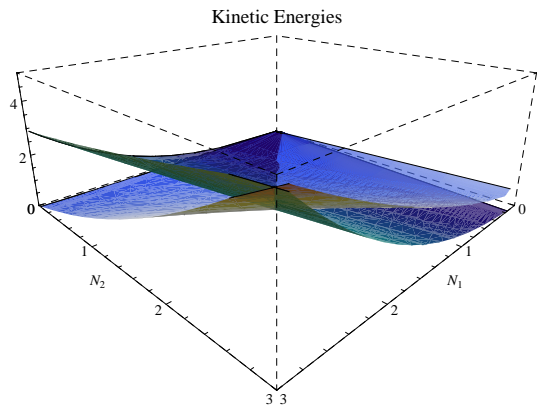}
\caption{\label{fig:fig6}Kinetic energy for each separate state $K_1$,
  $K_2$, in a ground-first configuration in terms of $\mathcal{N}_1$
  and $\mathcal{N}_2$. The systems with $K_1 >K_2$ correspond to
  $\eta_2 < 1$ ($\mathcal{N}_1 > \mathcal{N}_2$), whereas those with
  $K_1 <K_2$ correspond to $\eta_2 > 1$ ($\mathcal{N}_1 <
  \mathcal{N}_2$).}
\end{figure}

\section{Stability of mixed states under small perturbations }
\label{sec:stability}
In order to promote the existence of mixed states beyond the
mathematical context, it is first necessary to prove their
stability. It is known that when all the particles are in the ground
or in an excited state, the configuration is stable under small radial
perturbations that strictly conserve the number of particles, 
$\delta N = 0$ \cite{Lee:1995af,Guzman:2004wj}. 

Instead, if the system is considered open, so that $\delta N = 0$ is
not demanded, an excited state emits particles, loses its nodes, and
eventually settles down on to a ground
state\cite{Guzman:2004wj,Guzman:2006yc}. On the contrary, ground
states can tolerate perturbations for which $\delta N \ne
0$\cite{Bernal:2006it}, and it is in this sense that ground states are
said to be stable, whereas excited states are unstable.
 
It is then expected that the excited state of a ground-first
configuration should remain unstable if there are very few particles
in the ground state. However, the results in the previous section show
that the properties of excited states change depending on how many
particles populate the ground state. 

This is actually the case of stability. We found that, by adding
particles to the ground state, it is possible to construct
ground-first configurations for which, under open conditions, the
mixed state is stable under radial perturbations. As particles do not
interact directly one with each other, the change in the stability of
the excited state is produced just by the gravitational interaction
with the ground state. 

Stability studies of single configurations have been done with perturbation 
theory and full numerical evolutions. In principle, the stability of mixed 
states can be also studied perturbatively (a work that is worth a separate 
manuscript). Instead, we have chosen to evolve numerically mixed states 
not only to determine their stability, but also their late time
behavior. We have done this for several ground-first configurations
with different values of $\eta_2$.

\subsection{Numerical perturbation}
\label{subsec:numpert}
The evolution of the mixed states is done by solving the discretized
version of the time dependent SP system~(\ref{eq:SPspherical}), taking
as the initial data the functions of the different unperturbed states
$\Psi_n$, see Eq.~(\ref{eq:Psi}), constructed in the previous section. 
We consider no perturbations other than those introduced
by the finite differencing error in the numerical integration.

The procedure followed in the construction of the solutions to
Eqs.~(\ref{eq:SPspherical}) can be summarized as follows.

\begin{enumerate}
\item The numerical grid is populated with the initial
  data $$\Psi_n^N(t_0=0,k\Delta r)=\phi_n(k\Delta r)\, ,$$ where $t_0$
  is an arbitrary initial time that we choose to be zero, $\Delta r$
  is the spatial resolution of the grid, the super-index $N$ labels
  the discretized numerical solution, $k$ is an integer that labels 
  the grid and $n$ runs from $1\dots\mathcal{I}$, $\mathcal{I}$ the 
  number of populated states. 

\item The gravitational potential $U^N(0,k \Delta r)$ is obtained by
  introducing $\Psi_n^N$ in Eq.~(\ref{eq:SPspherical}).

\item Using the obtained gravitational potential $U^N$ in
  Eqs.~(\ref{eq:SPspherical}), each populated state of the system is
  leaped forward in time a step $\Delta t$ getting $\Psi_n^N(\Delta
  t,k\Delta r)$. 

\item Repeating $j-1$ times the steps 2 and 3 above we obtain
  $\Psi_n^N(j\Delta t,k\Delta r)$ and $U^N((j-1)\Delta t,k \Delta r)$. 
\end{enumerate}

Because the SP system is discretized in order to solve it numerically,
it is expected that the numerical solution $\Psi_n^N$ differs from the
equilibrium value $\Psi_n$ in Eq.~(\ref{eq:Psi}) by a numerical error
$\Delta \Psi_n$, i.e.,
\begin{equation}\label{perturbedstate}  
\Psi_n^N(j\Delta t, k\Delta r)=e^{-i\gamma_nj\Delta t}\phi_n(k\Delta r)
+\Delta \Psi_n(j\Delta t,k\Delta r)\,.
\end{equation}
We then say that the system is perturbed without considering an
explicit perturbation, because the numerical error $\Delta \Psi_n$, that 
comes from the discretization, is considered as the perturbation itself; 
in fact, $\Psi_n^N$ behaves like a perturbed $\Psi_n$ during the evolution. 

We have to stress that during the numerical evolution we allow the system 
to eject particles. With the implemented open boundary conditions in the 
code, the condition for which the number of particles has to be preserved 
is not maintained.

\subsection{Perturbing the ground state}\label{PGS}
In Ref.~\cite{Guzman:2004wj} the discretization error was used to perturb 
a single ground state in order to study its stability and it was shown 
that the numerical evolution reproduces the results obtained with 
perturbation theory. Here we give a brief account of this result.

Using perturbation theory to first order, when a single ground state 
function $\Psi_1=\phi_1(r)e^{-i\gamma_1t}$ is perturbed with a small 
radial perturbation $\delta \Psi_1$, keeping the number of particles
constant, a new oscillation mode $\sigma$ appears in the perturbed
system $\Psi_{pert}=\Psi_1+\delta \Psi_1$. It is found that $\delta
\Psi_1$ is regular, spatially localized, and has an harmonic time
dependence (such time-dependence involves not only the new oscillation
mode $\sigma$, but also the eigenvalue $\gamma_1$). The system is then
said to be stable under small radial perturbations.

A quantity that gives relevant information about the new 
oscillation mode is the perturbed density. The ground state
non-perturbed density $\rho_1=|\Psi_1|^2$ is time independent, whereas
the perturbed one $\rho_{pert}=|\Psi_{pert}|^2$ has an harmonic time
dependence and oscillates with an angular frequency $2\pi \sigma$. To
first order in $\delta \Psi_1$, the perturbed density is
\begin{equation}\label{perturbed-density}
  \rho_{pert} = |\Psi_1+\delta \Psi|^2 = \rho_1(r) + \delta
  \rho(r)\cos(2\pi \sigma t) \, .
\end{equation}

These perturbed ground state properties are also found when the
ground state is evolved numerically without considering any explicit
perturbation, except for that inherent to the discretization. Using as
initial data the time independent function $\phi_1$ of a ground state, its  
temporal behavior is obtained as explained in the previous subsection.
A quantity that is monitored throughout the evolution is the function 
$\mathrm{Re}[\Psi_1^N(0,t)]$, which behaves harmonically in time as
expected. Its Fourier Transform (FT) shows a main harmonic mode that
matches the eigenfrequency $\gamma_1$ of the unperturbed ground 
state.

The density $\rho_1^N(t,0)=|\Psi_1^N(t,0)|^2$ also shows an harmonic
behavior in agreement with the perturbative
result~(\ref{perturbed-density}). It oscillates around the value of
the non-perturbed density $\rho_1$, and from its FT we observe that
its main oscillation mode coincides with $\sigma$. Finally, even with
the open boundary conditions in the numerical code, the number of
particles does not change, which is consistent with the results of
perturbation theory.

\begin{figure}[t]
\centering
\includegraphics*[width=7cm,angle=-90]{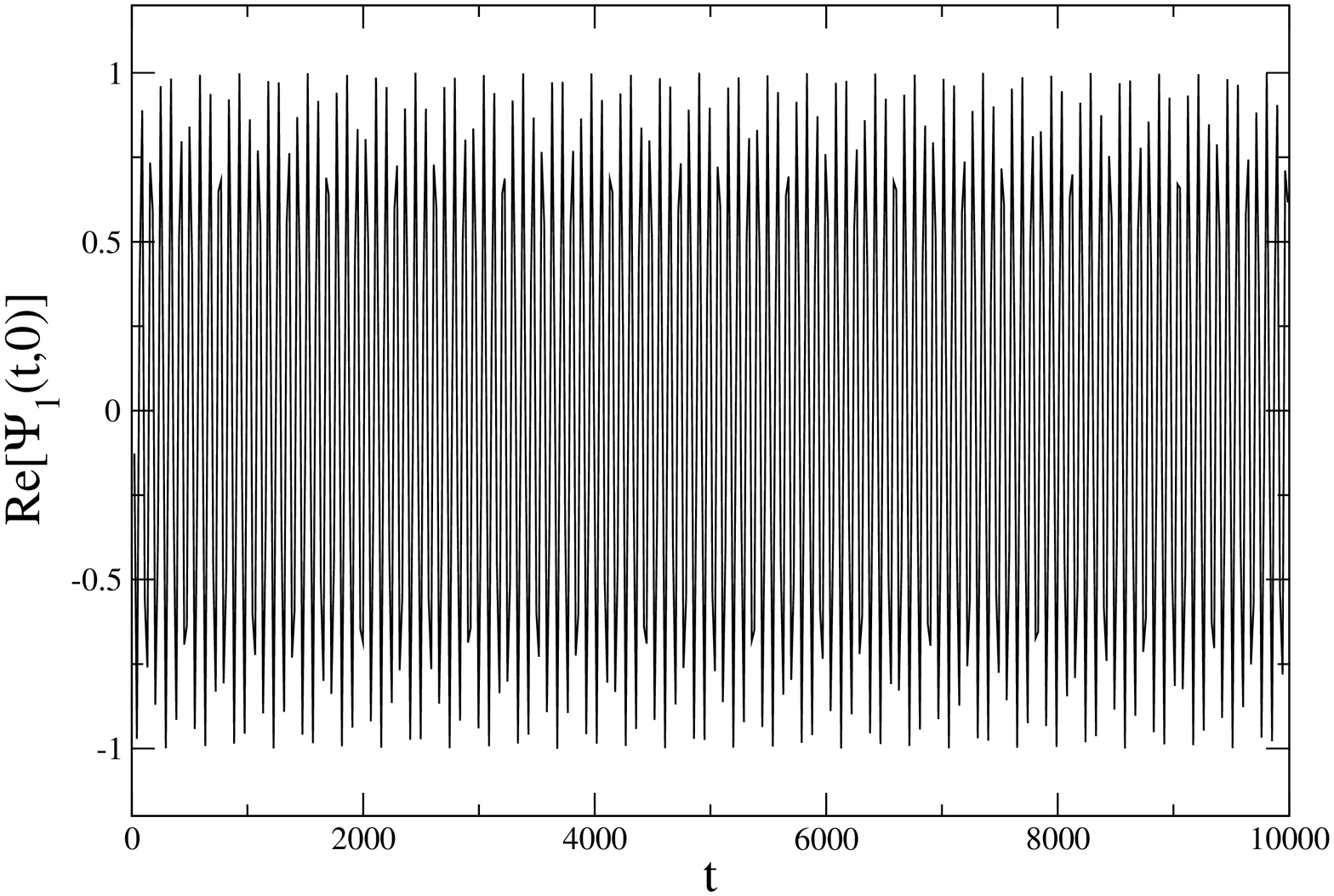}
\includegraphics*[width=7cm,angle=-90]{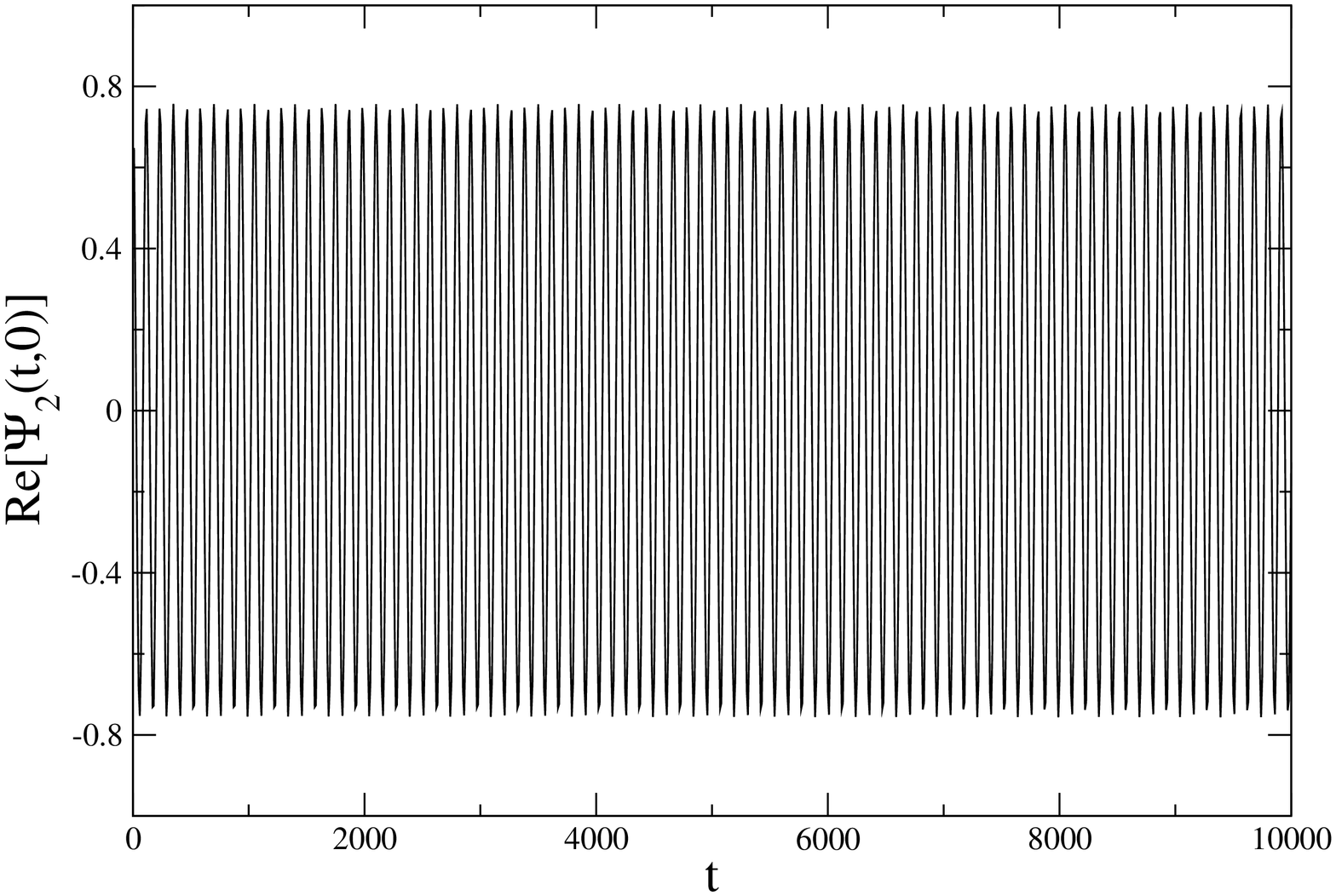}
\caption{Evolution of $\textrm{Re}[\Psi_1^N(t,0)]$ (top) and
  $\textrm{Re}[\Psi_2^N(t,0)]$ (bottom) for the mixed state
  $|N_1,\eta_2 \rangle$. An harmonic behavior is observed, and the 
  oscillation modes can be read from the corresponding FFT shown in
  Fig. \ref{fig3}.}\label{fig2}
\end{figure}

\begin{figure}[htp]
\includegraphics*[width=7cm,angle=-90]{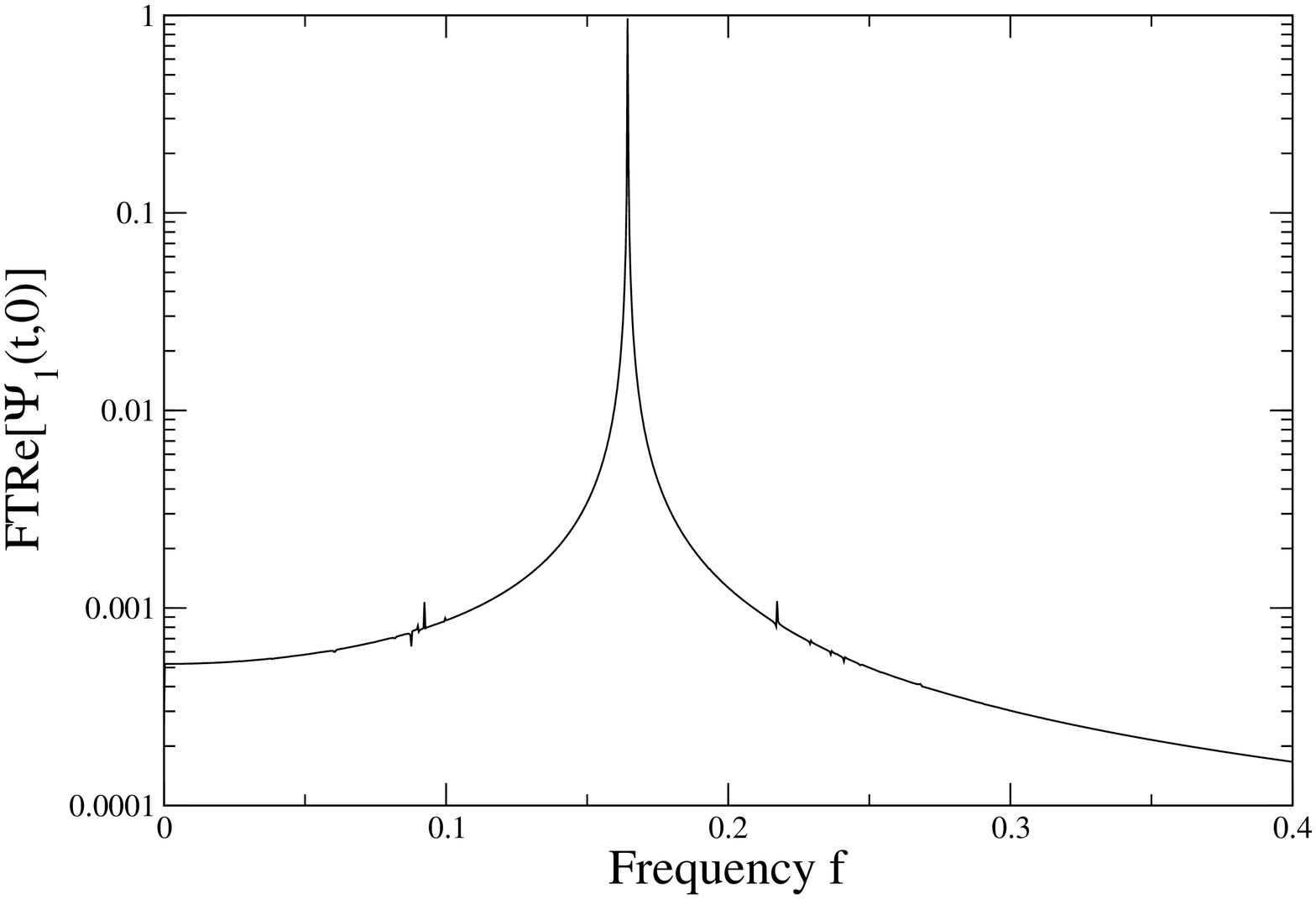}
\includegraphics*[width=7cm,angle=-90]{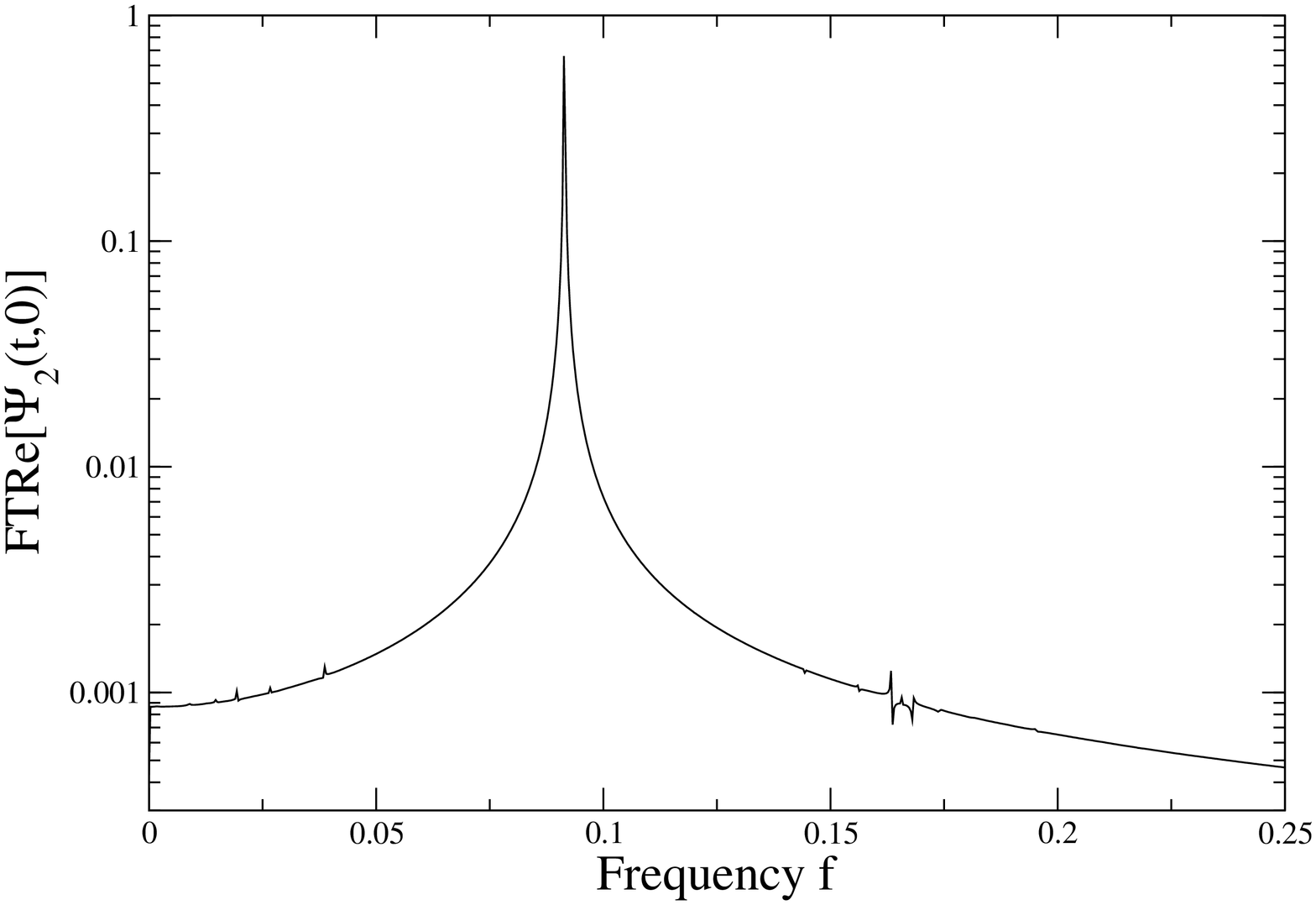}
\caption{\label{fig3}(Top) FFT of $\textrm{Re}[\Psi_1^N(t,0)]$ for
  the $|N_1,\eta_2 \rangle$ mixed state. The main mode at $f = 0.164$
  corresponds to the angular frequency $2\pi f= 1.033$ which
  coincides, in good approximation, with the value of $\gamma_1$ of
  the unperturbed wave function $\Psi_1$, see
  Table~\ref{table1}. (Bottom) FFT of $\textrm{Re}[\Psi_2^N(t,0)]$
  for the $|N_1,\eta_2 \rangle$ mixed state; the main mode at $f =
  0.091$ corresponds to the angular frequency $2\pi f= 0.574$, which
  also coincides with the value of $\gamma_2$ of the unperturbed wave
  function $\Psi_2$, see Table \ref{table1}.}
\end{figure}

\subsection{Perturbing stable mixed states}
In order to study the stability of a mixed state, we evolve it
following the steps described in \ref{subsec:numpert}. The general
idea is to compare the behavior of $\mathrm{Re}[\Psi_{n}^N(t,0)]$,
$\rho_n^N(t,0)$, and the number of particles $\mathcal{N}_n^N(t)$ for
each occupied state with the behavior of the corresponding quantities for 
a single ground state. We can conclude that there is evidence of the
stability of the mixed configuration if those behaviors are similar.

In fig.~\ref{fig2}, we show the numerical evolution of
$\mathrm{Re}[\Psi_{n}^N(t,0)]$ for a ground-first configuration with
$\eta_2 =1.1$, i.e., a $|\mathcal{N}_1,1.1\rangle$ state. The wave
functions behave harmonically, and their FTs, presented in
Fig.~\ref{fig3}, show that the main harmonic mode of each $\Psi_{n}^N$
corresponds to the angular frequency $\gamma_n$ of the unperturbed
$\Psi_n$.

\begin{figure}[htp]
\includegraphics*[width=7cm,angle=-90]{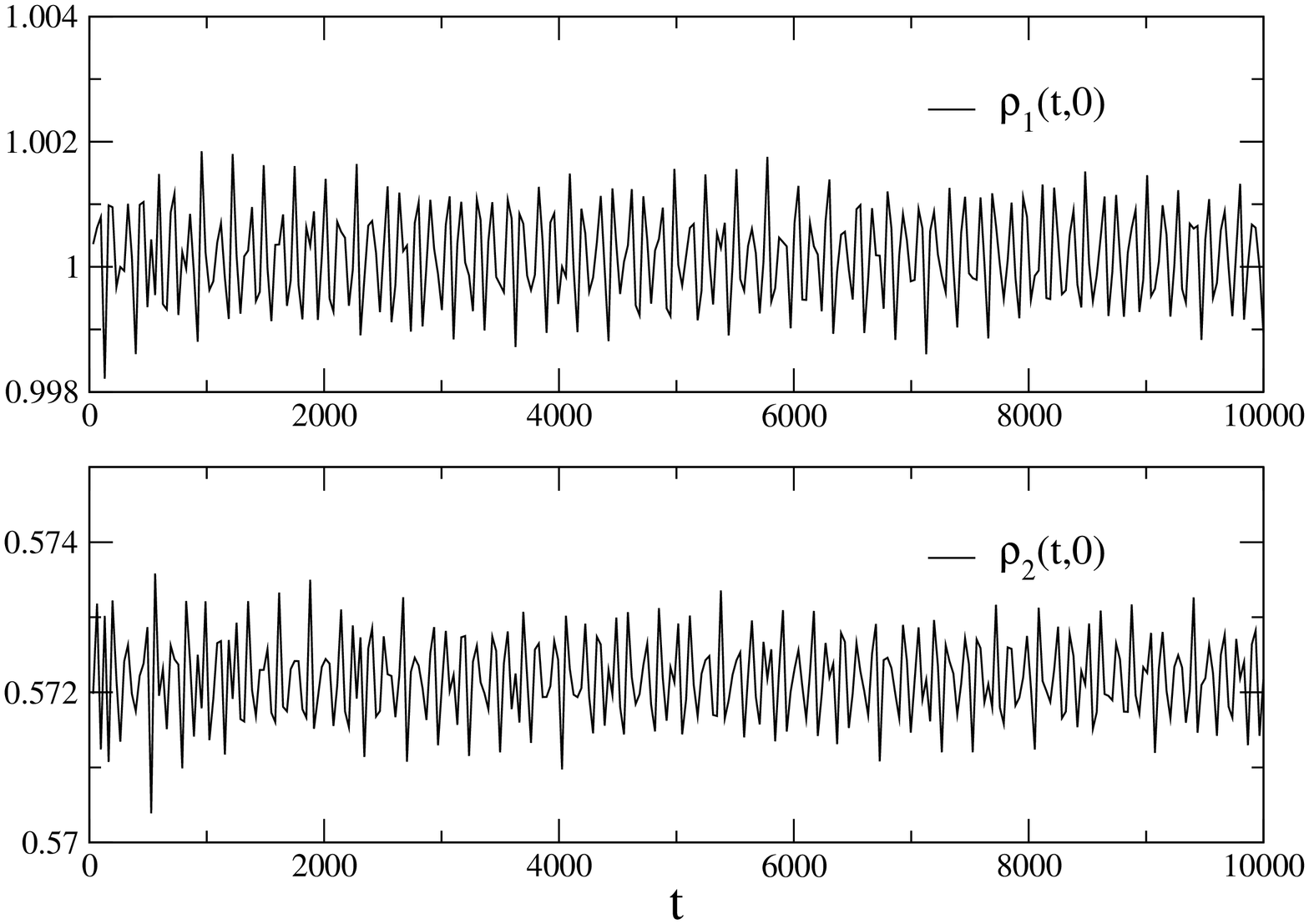}
\includegraphics*[width=7cm,angle=-90]{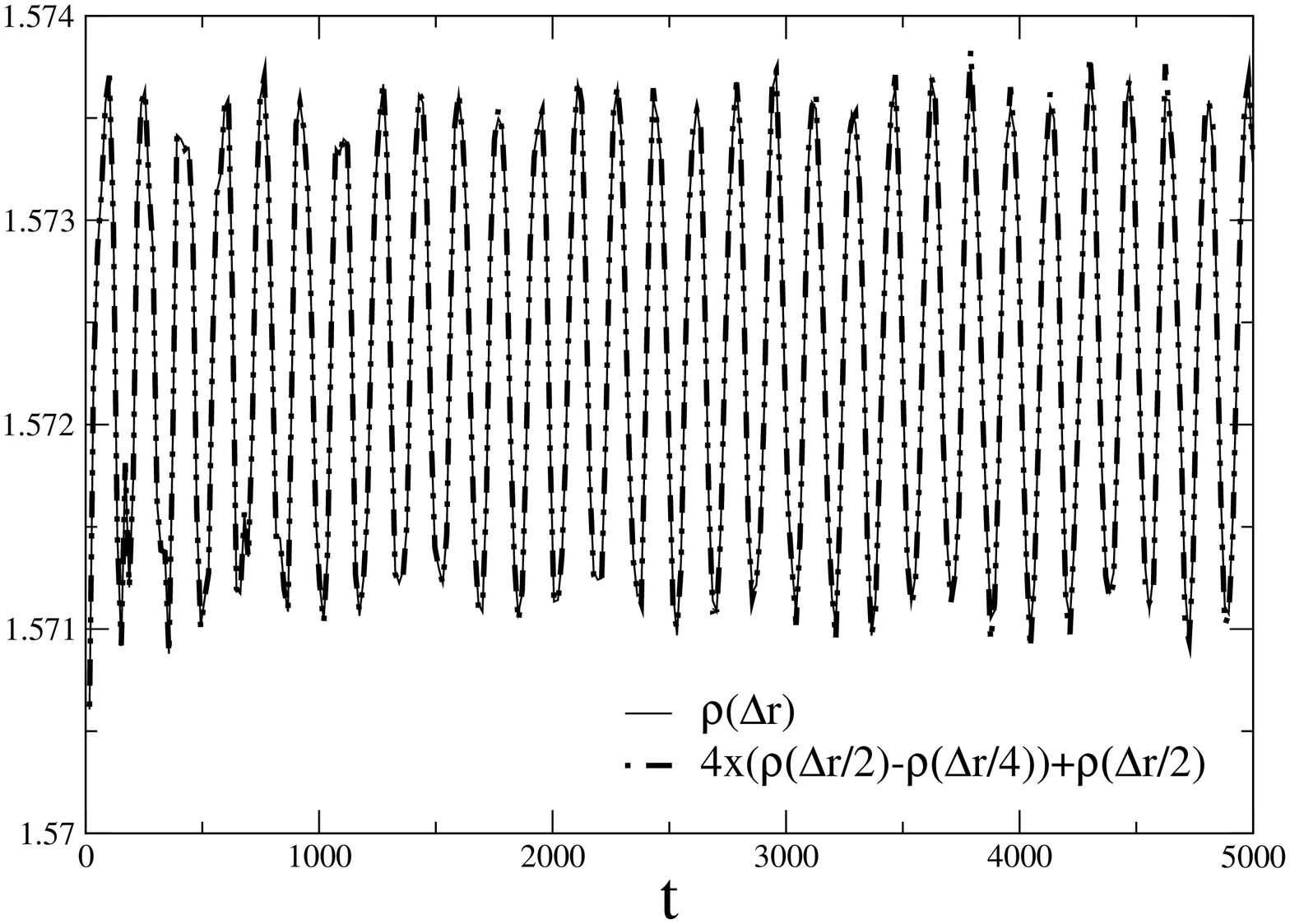}
\caption{ (Top) The perturbed central densities $\rho_n^N=|\Psi_n^N(t,0)|^2$ 
for the $|N_1,\eta_2 \rangle$ configuration are shown. They oscillate
around the constant density values of the unperturbed states,
corresponding to $\rho_1=1.0$ and $\rho_2=0.485$. Such oscillatory
behavior is the expected one for stable states (see Sec.~\ref{PGS} for
details). (Bottom) Overlap between $\rho^N(\Delta r)$ and
$4(\rho^N(\Delta r/2) - \rho^N(\Delta r/4))+\rho^N(\Delta r/2)$. This plot shows that the numerical
evolution is second order convergent.
}
\label{fig4}
\end{figure}

Furthermore, in Fig.~\ref{fig4} we show the numerical values of the 
central densities $\rho_n^N=|\Psi_{n}^N(t,0)|^2$, and it is clear that
they oscillate closely around the value of the unperturbed densities
(which are formally time-independent), $\rho_1=|\phi_1|^2=1.0$ and 
$\rho_2=|\phi_2|^2=0.572$.

\begin{figure}[htp]
\includegraphics*[width=7cm,angle=-90]{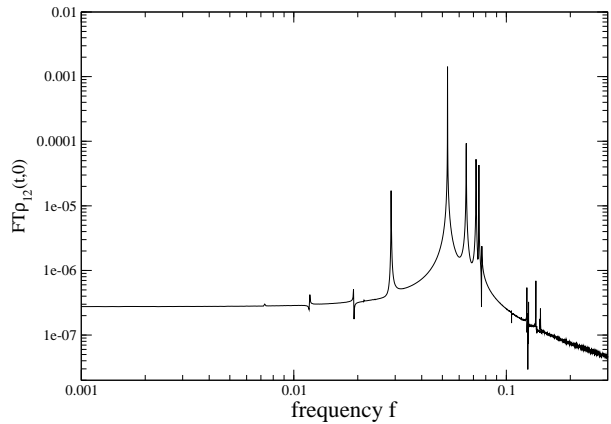}
\caption{FFT of the numerically-perturbed total density
  $\rho^N(t,0)$. The main mode appears at $f =0.065$, which should
  correspond to the quasinormal mode of oscillation of the perturbed
  configuration.}
\label{fig5}
\end{figure}

The bounded oscillations suggest that the numerical perturbations of
each $\Psi_n$ are spatially localized and have an harmonic time
dependence. In the bottom panel of Fig.~\ref{fig4}, we show the
central value of the total density $\rho^N=\rho_1^N+\rho_2^N$, with
its corresponding FT in Fig.~\ref{fig5}. As in the case of a single
ground state, for which its quasinormal mode is the mean harmonic mode
of the perturbed density, we expect that the mean harmonic mode shown
in this figure corresponds to the characteristic oscillation mode of
the perturbed state $|\mathcal{N}_1,1.1\rangle$.

\begin{figure}[htp]
\includegraphics*[width=7cm,angle=-90]{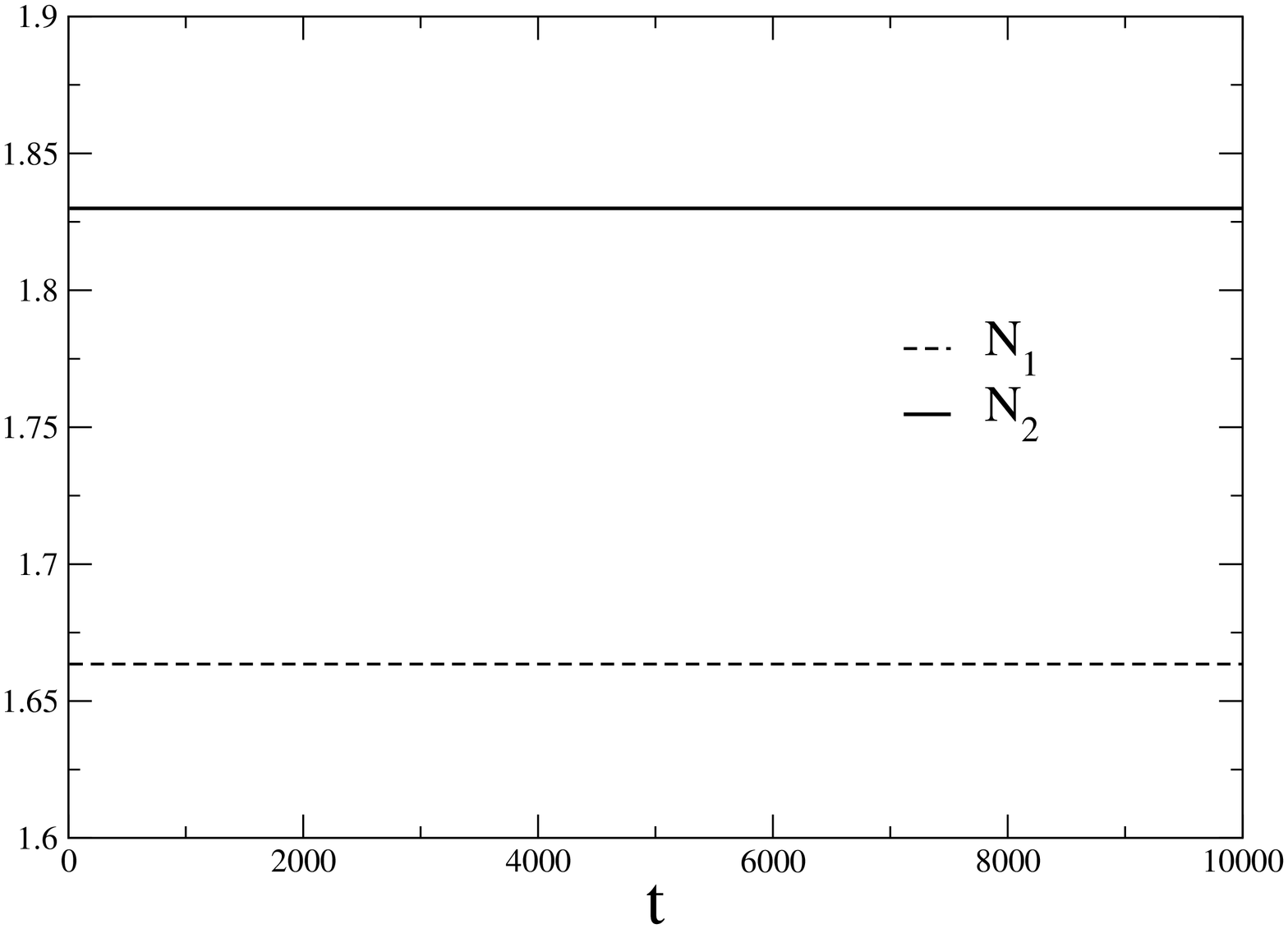}
\includegraphics*[width=7cm,angle=-90]{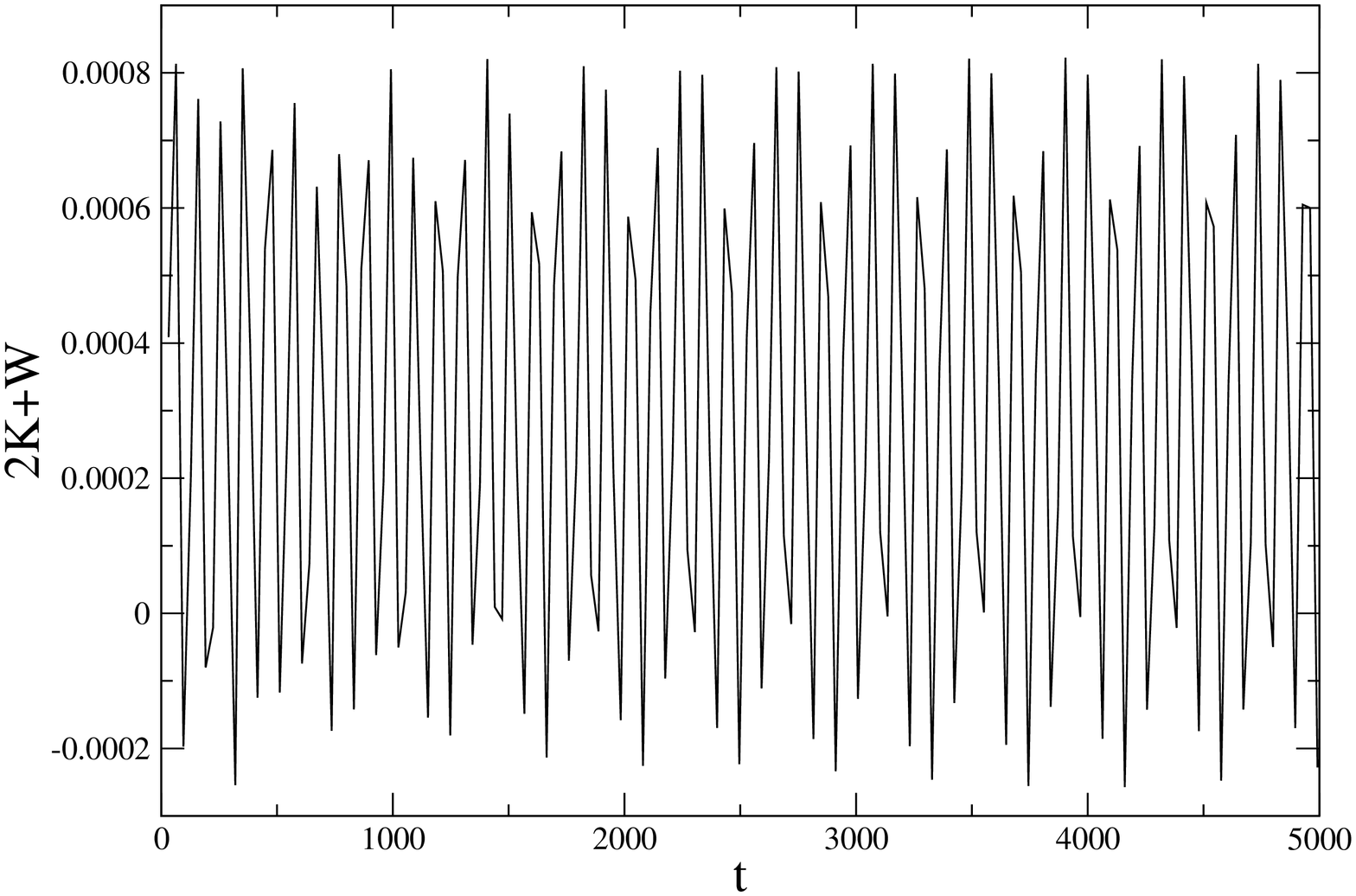}
\caption{(Left) Number of particles for each state of the mixed
  configuration $|N_1,\eta_2 \rangle$ is shown. It is observed that
  the number of particles is conserved all along the numerical
  evolution. (Right) The value $2K+W$ is shown. Since it 
oscillates around zero, it is inferred that the mixed configuration is a 
virialized system}\label{fig6}
\end{figure}

In Fig.~\ref{fig6}, we verify that the number of particles for each 
occupied state is conserved separately, so that the same happens for
the total number of particles. In the bottom panel of Fig.~\ref{fig6},
the relation $2K+W$ for the system is presented, which in turn shows
that the configuration remains virialized during the evolution.

\subsection{Perturbing unstable mixed
  states \label{sec:pert-unst-mixed}}
We performed the evolution for several ground-first states with
$\phi_1(0)=1.0$ and different values of $\eta_2$. A general result is
that if $\eta_2 <1.1$ the mixed state behaves similarly as the state
$|N_1,1.1 \rangle$. That is, each state of the configuration evolves
harmonically with the main angular frequency $\gamma_n$ of the
unperturbed wave function $\Psi_n$, the oscillations of the central
densities $\rho_n^N$ are bounded, and the number of particles in each
state is conserved. We take for granted that these characteristics are
evidence of the stability of these ground-first configurations.

On the other hand, the evolution of configurations with $\eta_2 > 1.1$
differ considerably. In the top panel of Fig.~\ref{fig8} we show the
early evolution of the central densities $\rho_n^N$ for a ground-first
configuration with $\phi_1(0) =1.0$ and $\eta_2 =1.6$, which is the
state $|N_1,1.6 \rangle$.
 
At the beginning, they just oscillate around the values of the
unperturbed initial state, $\rho_1(0) =1.0$ and $\rho_2(0)
=0.873$. However, after a while the amplitudes of the oscillations
start to grow. In fact there is a period of time in which it is possible to 
fit each $\rho_n^N$ with a function of the form
\begin{equation}
  \label{eq:unstable}
  \rho_n \simeq a_1 e^{a_2 t} \cos(a_3 t + a_4) \, , 
\end{equation}
where the $a$'s are constants. This behavior suggest that the
perturbations of the states $\Psi_n$, besides their harmonic time
dependence, are exponentially unstable since $a_2 > 0$ in general.

\begin{figure}[htp]
\includegraphics*[width=7cm,angle=-90]{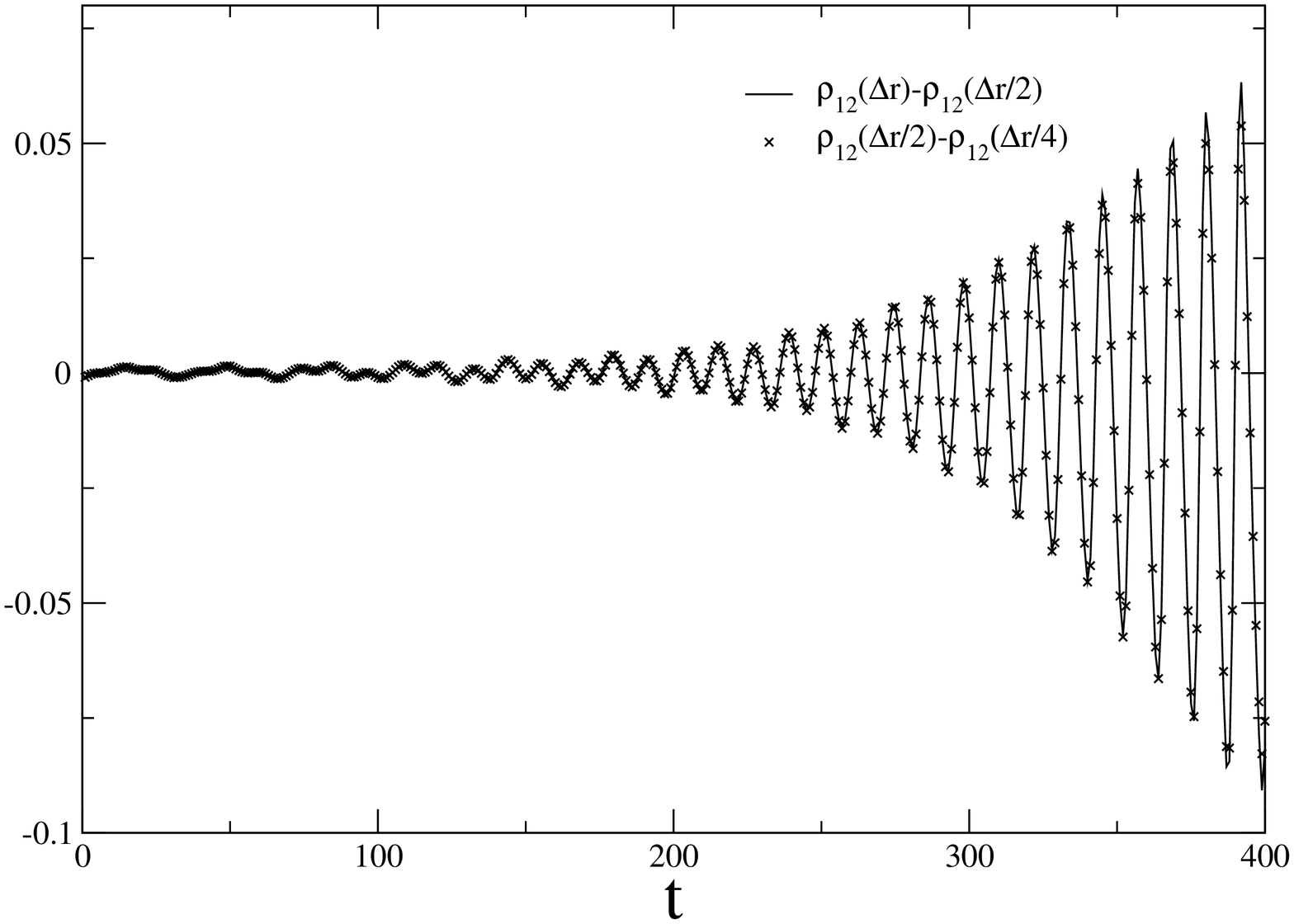}
\includegraphics*[width=7cm,angle=-90]{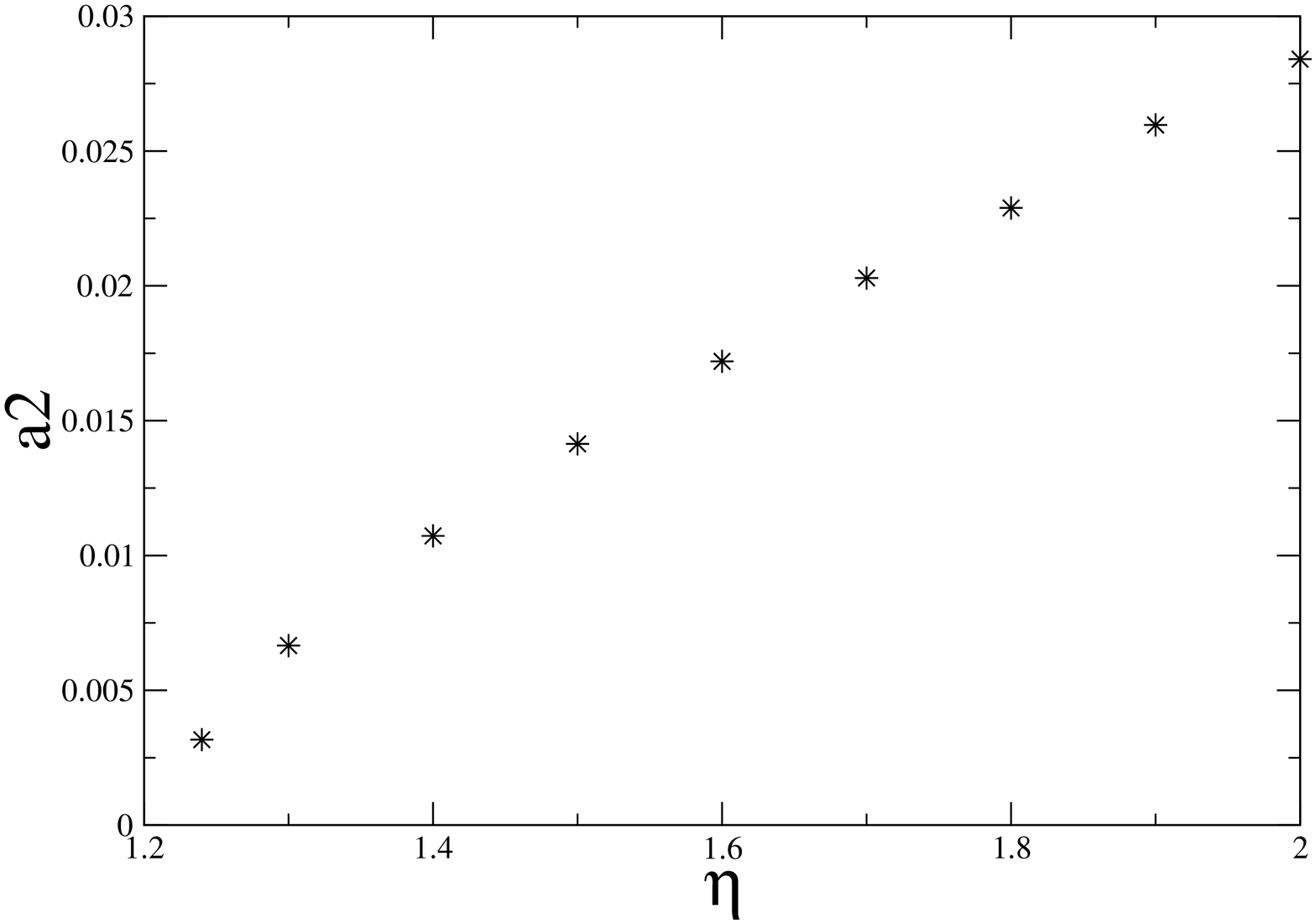}
\caption{(Top) The perturbed central density $\rho_2^N$ of a unstable
  configuration with $\eta=1.6$ is shown. The overlap between
  $\rho_2^N(\Delta r)-\rho_2^N(\Delta r/4)$ and $4(\rho_2^N(\Delta
  r/2)-\rho_2^N(\Delta r/4))$ show that $\rho_2^N$ converges at second order.  
(Bottom) Coefficient $a_2$ of the exponential
function~(\ref{eq:unstable}) used to fit the growth in the
oscillations of the $\rho_2^N$ for different unstable configurations
are shown. It is null for the threshold value of $\eta_2 \simeq 1.13$,
see text for details.}\label{fig7}
\end{figure}

From Fig.~\ref{fig8}, it is possible to infer that the exponential
growth of the density amplitude of the excited state, which starts
almost from the very beginning of the evolution, acts as a trigger for
the instability of the complete configuration. For completeness, we
show in the top panel of Fig.~\ref{fig7} the convergence of
$\rho_{12}^N(t,0)$ for different spatial resolutions. We can conclude
that the exponential growth of perturbations is a physical
characteristic and not a spurious numerical result. 

Moreover, the number of particles in each state of the configuration
$|N_1,1.6 \rangle$ is not conserved. In Fig.~\ref{fig10} we show that
both the ground and the first excited states lose particles. Even
more, the behavior of each $\mathrm{Re}[\Psi_{n}^N(t,0)]$ also shows
that their angular frequency change with time, and that only at the
early stages of the evolution the frequencies coincide with the
unperturbed values $\gamma_1=-1.163$ and $\gamma_2=-0.677$, see the
top panel of Fig.~\ref{fig9} (the way in which the plotted quantities
are computed is explained in the next section). Therefore, the
configuration $|N_1,1.6 \rangle$ is not stable.

We performed the evolution for several configurations with 
different values of $\eta_2 > 1.2$, and we noticed that the 
speed of growth of the perturbation amplitudes increases for larger
values of $\eta_2$. In order to quantify the speed of growth in terms of 
$\eta_2$, during the early times of the evolution we fit the central
density of the excited state $\rho_2^N$ with a function of the
form~(\ref{eq:unstable}). 

The values of coefficient $a_2$ for the exponential coefficient in
terms of $\eta_2$ are plotted in Fig.~\ref{fig6}. A linear
extrapolation indicates that $a_2 \to 0$ as $\eta_2 \to 1.13$. As we
already found that configurations with $\eta_2 < 1.1$ do not exhibit
exponential growth, then we take $\eta^{\mathrm{tresh}}_2 \sim 1.13$ as
a threshold value that separates stable and unstable configurations.
 
This result has been obtained from configurations with $\phi_1(0)=1.0$
and different values of $\eta_2$. Since mixed configurations obey the
scaling symmetry~(\ref{eq:escalingsol}), for which $\eta_2$ is an
invariant quantity, then this threshold value should hold for all
configurations.

\section{Late time behavior of unstable mixed states}
In this section we present evidence that shows that unstable
ground-first configurations, when perturbed, evolve towards a stable
configuration, in a process that parallels the evolution of single
unstable states into the ground one, see\cite{Guzman:2006yc}.

During the stabilization process, we will see that the excited state
loses particles and the node in its radial profile
disappears. Likewise, the ground state loses particles as well, but a
node appears in its radial profile. In any case, it will be possible
to identify the final stable configuration the system is settle down
onto.

A useful procedure is to follow the evolution of the effective
eigenfrequency defined by
\begin{equation}
  \label{eq:gamma}
  \gamma_i^N \equiv \frac{1}{\int |\Psi_i^N|^2 dv} \left(
    -\frac{1}{2} \int \Psi_i^N \nabla^2 \Psi_i^N dv + \int U|\Psi_i^N|^2
    dv \right) \, ,  
\end{equation}
which is a generalization of the eigenfrequency obtained from the
Schroedinger equation~(\ref{eq:SPspherical}) in the case of stationary
solutions of the form~(\ref{eq:Psi}), i.e.,
\begin{equation}
  \label{eq:gamma}
  \gamma_i = \frac{1}{\int \phi_n^2 dv} \left(
    -\frac{1}{2} \int \phi_i \nabla^2 \phi_i dv + \int U|\phi_i^2 dv
  \right) \, .
\end{equation}

The precise identification of the final configuration is a tricky
task, as the ground and the excited state interchange the node in
their radial profile during the evolution. We have found that the
labeling of the states, whether ground or first excited state, depends
basically in the relative values of $\gamma_1^N$ and $\gamma_2^N$. As
in the case of stationary solutions, we shall call ground state the
state with the lowest value of $\gamma^N$.

\begin{figure}[htp]
\includegraphics*[width=7.5cm,angle=-90]{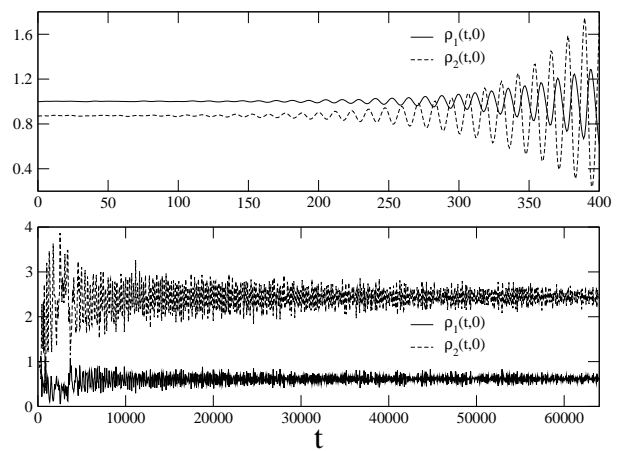}
\caption{Evolution of $\rho_i^N(t,0)$ for a unstable ground-first
  configuration with $\eta=1.6$. (Top) Early time behavior; (Bottom)
  late time behavior. The system eventually settles down onto a
  stable, stationary, configuration.}\label{fig8}
\end{figure}

\begin{figure}[htp]
\includegraphics*[width=7.5cm,angle=-90]{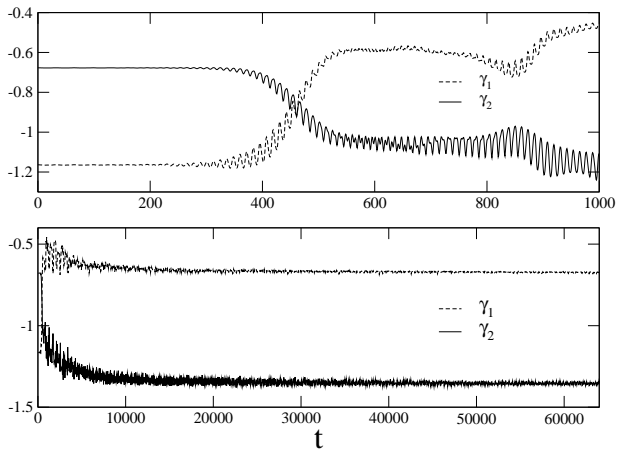}
\caption{Evolution of $\gamma_i^N$ for a unstable ground-first configura
tion with $\eta=1.6$. (Top) Early time behavior; (Bottom) late time
behavior, see text below. }\label{fig9}
\end{figure}

Now, we are going to describe the main stages in the evolution of the
ground-first configuration $|N_1,1.6\rangle$ in terms of
$\rho_i^N(t,0)$, $\gamma_i^N(t)$, and $\mathcal{N}_i^N(t)$. At the
beginning of the evolution, $\rho_i^N(t,0)$ oscillate with small
amplitude around the unperturbed values $\rho_1=1.0$ and
$\rho_2=0.873$, see the top panel in Fig.~(\ref{fig8}). The same
behavior is found for $\gamma_i^N(t)$, which oscillate around
$\gamma_1=-1.163$ and $\gamma_2=-0.677$, see the top panel in
Fig.~\ref{fig9}. 

\begin{figure}[htp]
\includegraphics*[width=7cm,angle=-90]{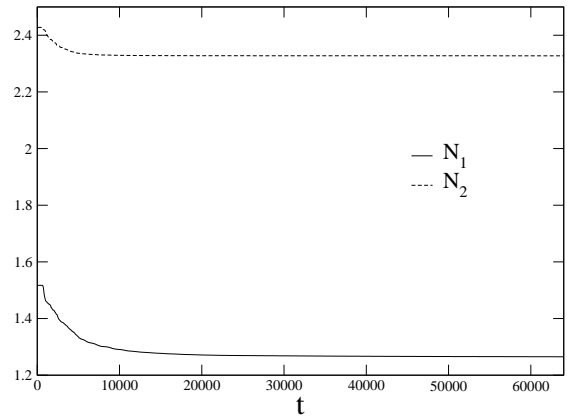}
\caption{Evolution of $N_1$ and $N_2$ for a unstable ground-first
  configuration with $\eta=1.6$.} \label{fig10}
\end{figure}

During this stage $\gamma_1^N < \gamma_2^N$, we can clearly see that
the radial profile of $\textrm{Re}[ \Psi_1^N (t,0)]$ shows no nodes,
whereas $\textrm{Re}[ \Psi_1^N (t,0)]$ has one node. The number of
particles $\mathcal{N}_i^N$ in each state remains constant, see
Fig.~\ref{fig10}.
 
Later on, the amplitude of each $\rho_i^N(t,0)$ starts to grow
exponentially as discussed before, see for instance
Fig.~\ref{fig8}. Meanwhile, $\gamma_i^N$ oscillate with bigger amplitudes   
and move away from their initial values, in such a way that $\gamma_2^N$ 
decreases and $\gamma_1^N$ increases, see Fig.~(\ref{fig9}). At the
same time, the number of nodes in the radial profiles
$\mathrm{Re}[\Psi_i^N(t,0)]$ changes quickly and the number of
particles of both states starts to decrease, see Fig.~(\ref{fig10}).
 
Then the system enters in a stage in which it experiences the most
violent changes. The amplitudes of the oscillations of $\rho_i^N(t,0)$
and $\gamma_i^N(t)$ becomes larger (Figs.~(\ref{fig8})
and~(\ref{fig9})), and the values of $\gamma_i^N(t)$ intersect and
continue moving away from each other. The number of nodes in the
radial profiles continues changing, and the number of particles of
each state decreases quickly, see Fig.~(\ref{fig10}).
  
Finally, the system relaxes and the values of $\rho_i^N(t,0)$ and
$\gamma_i^N(t)$ start to converge and oscillate around a fixed
value. It can be noticed that $\gamma_1^N > \gamma_2^N$, and that the
radial profiles have interchanged nodes. Also the number of particles
in each state stabilizes around fixed values. All this description
strongly suggests that the system is reaching a stable configuration.

In order to verify that the final configuration really corresponds to
a stable one, we read from the numerical results a set of parameters
that can help us to construct an equilibrium configuration as
described in Sec.~\ref{subsec:mixstatescons}. For the particular case
studied here, we find that $\rho_2^N=0.617$ (recall that the excited
state became the ground one), and $\eta_2 =
\mathcal{N}_1^N/\mathcal{N}_2^N =0.544$.

We then construct an equilibrium configuration with the aforementioned
values as input parameters, compute all relevant quantities, and
compare them with their (numerically) evolved counterparts. The
resulting values are shown and compared in Table~\ref{table2}. Based
on the fact that those values coincide in good approximation, we can
affirm that the unstable configuration $|N_1,1.6 \rangle$, evolves
towards a stable configuration with $\eta_2 =0.544$ and
$\phi_1(0)\approx 1.557$. 

\begin{table*}
\begin{ruledtabular}
\begin{tabular}{cccccc}
  & $\phi_1(0)$ & $\phi_2(0)$ & $\gamma_1$ & $\gamma_2$ & $U(0)$ \\
  \hline
  Equilibrium & 1.557 & 0.795 & -1.355 & -0.692 & -2.461 \\
  \hline 
  Final & 1.563 & 0.786 & -1.355 & -0.690 & -2.461
\end{tabular}
\end{ruledtabular}
\caption{Central values of the excited states
  $\phi_n(0)$, eigenvalues $\gamma_n$, kinetic $K$ and gravitational
  $W$ energies, and the total number of particles $\mathcal{N}$ of the
  mixed states. Because of the similarity between the final values of
  the evolved configuration shown in Figs.~\ref{fig8}-~\ref{fig11}, it
  can be concluded that the final state certainly corresponds to an
  equilibrium configuration.}
\label{table2}
\end{table*}

\begin{figure}[htp]
\includegraphics*[width=7cm,angle=-90]{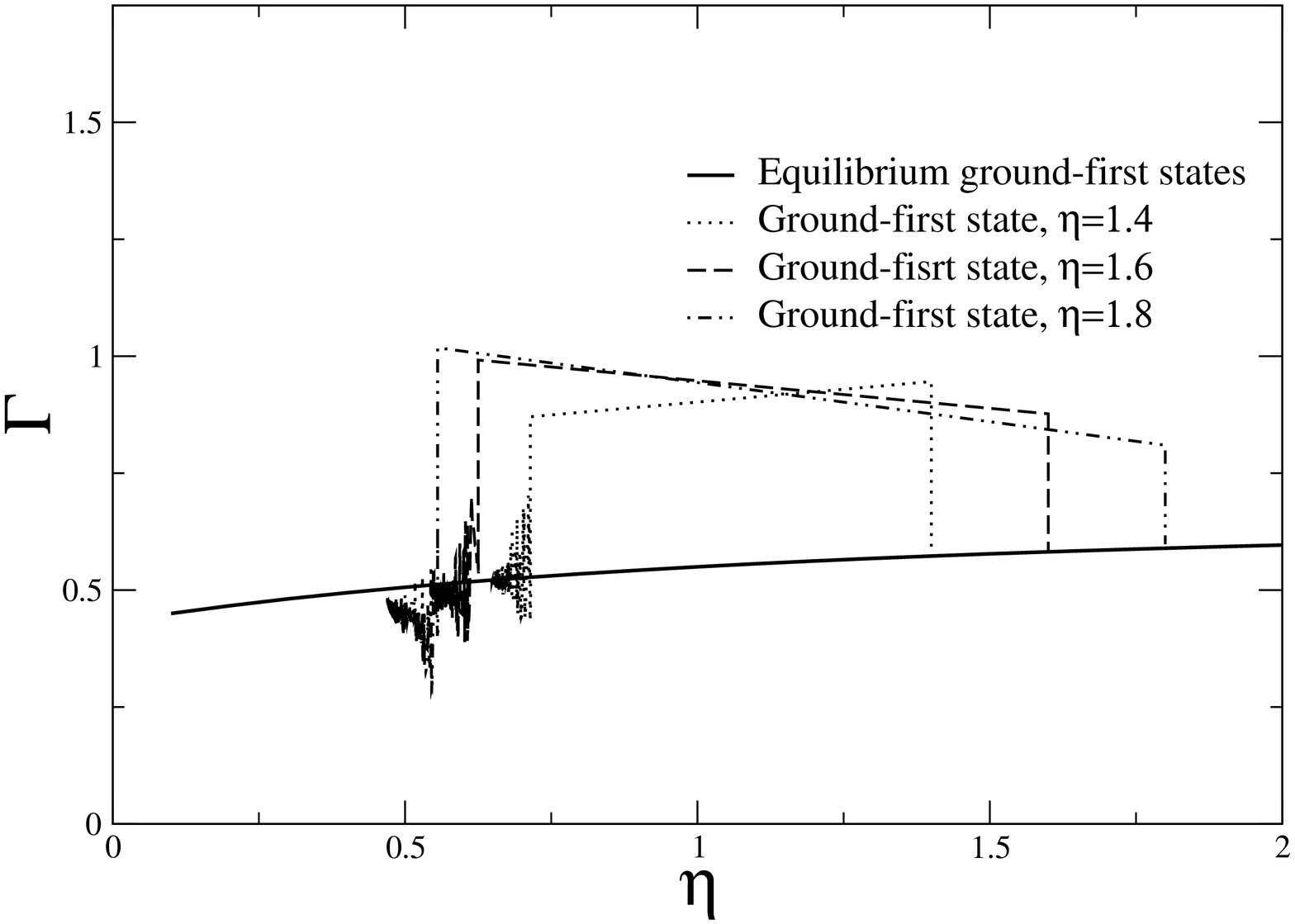}
\caption{Fate of unstable configurations with $\eta=1.4$,$\eta=1.6$
  and $\eta=1.8$, in terms of parameter $\Gamma$, see
  Eq.~(\ref{eq:ratio-freq}). Notice that the configuration oscillates
  around the line representing stationary equilibrium configurations
  on the stable branch $\eta_2 < 1.13$, see
  Sec.~\ref{sec:pert-unst-mixed}.}\label{fig11}
\end{figure}

Long evolutions for several configurations for which initially $\eta_2
> 1.1$ and $\phi_1(0)=1.0$, show that their behavior is very much like
the one observed in the unstable configuration with $\eta =1.6$. In
all cases, they evolve towards a stable configuration, as exemplified
in Fig.~(\ref{fig11}).

Here we have defined another scale-invariant quantity for the ratio
between the eigenfrequencies of the equilibrium configurations,
\begin{equation}
  \label{eq:ratio-freq}
  \Gamma_n \equiv \gamma_n /\gamma_1 \, ,
\end{equation}
such that again $\Gamma_1 = 1$. We can plot the resulting values of
ground-first stationary solutions on the plane $(\eta_2,\Gamma_2)$,
which are then represented by the solid line in Fig.~\ref{fig11}. We
notice that equilibrium configurations with initial $\eta_2 > 1.1$
evolve towards an equilibrium configuration with $\eta_2< 1.1$.
 
\section{Rotation Curves in scalar field galaxy halos}
\label{sec:RotCurves}
We mentioned before that our main motivation for studying mixed
configurations was the possibility that a scalar field would be the
dark matter in galaxies. For that, we wanted to explore the
capabilities of a scalar field to form realistic galaxy halos.

A crude estimation of rotation curves in mixed states was first
presented in Ref.\cite{Matos:2007zza}, under the (then untested)
assumption that they were stable, a feature we can now consider firmly
confirmed by the results of the present work.

For the mixed states studied in the previous sections, we can
calculate the velocity of test particles moving along circular orbits
in the gravitational potential sourced by the mixed states
configurations via the Newtonian formula
\begin{equation}
  v(r) = \sqrt{\mathcal{N}(t,r)/r} \, , \label{vel}
\end{equation}
where $\mathcal{N}(t,r)$ is the total number of particles inside the
radius $r$ obtained from the numerical equilibrium configurations of
the SP system~(\ref{eq:SPspherical}). 

\begin{figure}[htp]
\includegraphics*[width=8cm]{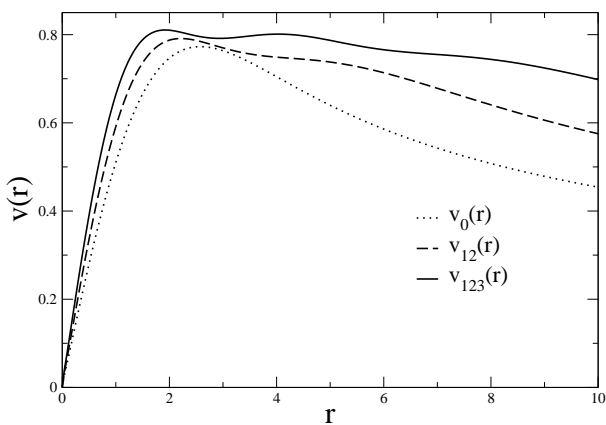}
\caption{\label{fig12} The rotation curve $v(r)$, see
  Eq.~(\ref{vel}), for the single ground state $|N_1=2.0622 \rangle$,
  and the mixed configurations $|N_1,1.1 \rangle$ and $|N_1, 0.96,
  0.91\rangle$, see also Table~\ref{table1}. Notice that the flatness
  of the curve is improved if more excited states are taken into
  account.}
\end{figure}

The results are shown in Fig.~\ref{fig12}. We can see a noticeable
improvement in the flatness of the rotation curve at large radii as
long as more excited states are taken into account. Some comments are
in turn.

The circular velocity of mixed configurations shows some flat profile
at intermediate radii, whereas the typical Keplerian tail shows up at
large radii, indicating that at the end we are dealing with a
localized object of finite size. We also note the existence of some
ripples in the velocity profile, which are a consequence of the nodes
present in the radial functions of excited states $\phi_n$. 

The height and position of the first peak in the velocity profile are
approximately set by the ground-state wave function $\phi_1$, whereas
the total size of the objects is fixed by the profile of the most
excited state.

At this point, we cannot say that mixed states are already strong
candidates to explain galaxy halos, but for that a more complete study
would be necessary, like the one pursued
in\cite{Sin:1992bg,Ji:1994xh,Lesgourgues:2002hk,Arbey:2003sj}, where
also baryons were included in the equations of motion.

We can though provide some clues about the possible physical features
of a scalar field galaxy halo. In most scalar field dark matter
models, the scalar field mass is usually very light, around $\mu \sim
10^{-23} \, \mathrm{eV}$; for such a small value, it was possible to
consider that single state configuration, whether the ground or any of
the excited ones, accounted for the complete halo configuration. This
was a central assumption in most of previous references about scalar field dark
matter
models\cite{Sin:1992bg,Sahni:1999qe,Matos:2000ss,Hu:2000ke,Lesgourgues:2002hk}.

The Compton wavelength of the scalar field is very large for usual
standards, because $\lambda_C = \mu^{-1} \simeq 10 \,
\mathrm{pc}$. In the case of a Newtonian configuration, the size of
the bounded object scales like $R = r/\lambda^2$, where $\lambda$ is
the scaling parameter in Eq.~(\ref{eq:escalingsol}), which in turn is
related to the central field value of the configuration through
$\lambda^2 = \phi_1(0)$.

Following previous works, the scaling parameter is estimated to be
$\lambda \sim 10^3$\cite{Guzman:2003kt,Guzman:2004wj}. This implies
that a stable single-state equilibrium configuration can model a
galaxy halo of a size of around $5-7$ kpc. However, larger galaxy
halos, as is the typical case, are out of the capabilities of single
state configurations. This limitation can be noticed, for instance, in
the fits done in\cite{Arbey:2003sj}.

Mixed states can alleviate this limitation. As the size of the
configuration is determined by the most excited state, the scalar
field halos can be as large as necessary, and this helps to fit better
the RC in real halos. 

Moreover, as already noticed in Ref.\cite{Matos:2007zza}, mixed states
provides us with  more free parameters to play with. The extra
parameters are the occupation numbers of the mixed state, namely
$N_1$, $N_2$, $N_3$, etc. Except for the limitations imposed by
stability, these values would be determined initially, before the
collapse of the scalar field configuration, by the local environment a
scalar halo could be subjected to during the cosmological evolution.

\section{Conclusions}
\label{Conclusions}
We have shown, for the first time, the existence of stable
many-particle states made of scalar field in the Newtonian regime;
such states are the generalization of the well-known boson stars that
have been exhaustively studied in the literature. The possibility of
mixed states was already suggested in the seminal paper of Ruffini \&
Bonazolla about boson stars, but their existence and properties had
not been studied before

Detailed instructions were provided for their construction and
classification, but more importantly is that we established simple and
sufficient criteria to determine their stability properties. Even
though Newtonian configurations obey a scaling relationship, it was
possible to define invariant parameters that allows us to follow their
evolution and determine their final fate.

Some remarks were given regarding the importance these results may have
for scalar field dark matter models in describing the properties of
galaxy halos. However, more work is needed to have a complete picture
of all possibilities offered by scalar fields and their
gravitationally bounded configurations. This is an objective we will
pursue in future work that we expect to report soon elsewhere.


\begin{acknowledgments}
We are grateful to Jose Socorro for his encouragement and support. We
also thank Juan Barranco, Ulises Nucamendi, and Olivier Sarbach for
helpful discussions, and Francisco S. Guzm\'an for sharing his
numerical code, which we modified to account for multi-state
configurations. This work was partially supported by CONACYT grant
56946, DAIP and PROMEP-UGTO-CA-3. AB acknowledges support from CONACYT
grant 47641, and the kind hospitality of the Departamento de F\'isica
of the Universidad de Guanajuato, for a postdoctoral stay during which
this work was initiated.
\end{acknowledgments}

\bibliography{GravBEC}
\end{document}